\documentclass[12pt,fleqn]{article}
\usepackage[cp866]{inputenc}
\usepackage[english]{babel}
\usepackage{makeidx}
\usepackage{latexsym,amsfonts,amssymb,amsmath,longtable,amsthm}
\input amssym.def
\usepackage{latexsym}
\usepackage{graphicx}
\usepackage[all]{xy}
\usepackage{amsfonts}
\usepackage{amsmath}
\usepackage{mathrsfs}
\usepackage{color}
\usepackage{ulem}
\usepackage[table]{xcolor}

\def\e{\varepsilon}
\def\ov{\boldsymbol{\omega}}

\def\D{{\mathcal{D}}}

\def\F{{\mathscr{F}}}

\def\N{\mathbb{N}}

\def\Ha{{\mathfrak{H}}}

\def\F{{\mathcal{F}}}

\def\bfeta{\boldsymbol{\eta}}

\def\int{\intop}

\def\n{{\bf n}}

\newcommand{\wotwo}{\mathop{\mathaccent23{W}^{1,2}}\nolimits}

\newcommand{\woq}{\mathop{\mathaccent23{W}^{k,q}}\nolimits}

\newcommand{\arctg}{\mathop{\rm arctg}}

\def\div{\hbox{\rm div}\,}
\def\curl{\hbox{\rm curl}\,}

\newtheorem{theo}{\bf Theorem}
\newtheorem{lem}{\bf Lemma}
\newtheorem{lemA}{\bf Corollary}
\newtheorem{lemr}{\bf Remark}

\newcommand{\pr}{{\bf Proof. }}

\def\Xint#1{\mathchoice
   {\XXint\displaystyle\textstyle{#1}}%
   {\XXint\textstyle\scriptstyle{#1}}%
   {\XXint\scriptstyle\scriptscriptstyle{#1}}%
   {\XXint\scriptscriptstyle\scriptscriptstyle{#1}}%
   \!\int}
\def\XXint#1#2#3{{\setbox0=\hbox{$#1{#2#3}{\int}$}
     \vcenter{\hbox{$#2#3$}}\kern-.5\wd0}}

\def\dashint{\Xint-}

\newcommand{\esssup}{\mathop{\rm ess\,sup}}

\newcommand{\R}{{\mathbb R}}
\newcommand{\const}{{\rm const}}
\newcommand{\dist}{\mathop{\rm dist}}

\newcommand{\meas}{\mathop{\rm meas}}

\newcommand{\ve}{{\mathbf v}}
\newcommand{\ue}{{\mathbf u}}
\newcommand{\fe}{{\mathbf f}}
\newcommand{\diam}{\mathop{\rm diam}}
\renewcommand{\div}{\mathop{\rm div}}

\newcommand{\loc}{{\rm loc}}

\begin{document}
\title{Solution of  Leray's problem  for
stationary Navier-Stokes equations in
plane and axially symmetric spatial domains\footnote{
bounded domains,  axially symmetric domains,  stationary Navier--Stokes
equations, boundary--value problem.}}
\author{ Mikhail V. Korobkov\footnote{Sobolev Institute of Mathematics, Acad. Koptyug pr. 4, and Novosibirsk State University, Pirogova str., 2,
630090 Novosibirsk, Russia; korob@math.nsc.ru},  Konstantin
Pileckas\footnote{Faculty of Mathematics and Informatics, Vilnius
University, Naugarduko Str., 24, Vilnius, 03225  Lithuania;
konstantinas.pileckas@mif.vu.lt} \, and Remigio Russo\footnote{   Department of Mathematics and Physics, Second University of Naples, Italy; remigio.russo@unina2.it}$\;$}

\maketitle

\begin{abstract} We study the nonhomogeneous boundary value problem
for the Navier--Stokes equations of steady motion of a viscous
incompressible fluid in  arbitrary bounded multiply connected
plane or axially-symmetric spatial domains. We prove that this
problem has a solution under the sole  necessary condition of zero total flux through the boundary.
The problem was  formulated  by Jean Leray 80 years ago.
The proof of the main result uses Bernoulli's law
for a weak solution  to the Euler equations.
\end{abstract}

\bigskip
\setcounter{section}{0}
\section{Introduction}
Let
\begin{equation}\label{domain}
\Omega=\Omega_0\setminus
\bigl(\bigcup\limits_{j=1}^N\bar\Omega_j\bigr),  \ \ \bar\Omega_j\subset\Omega_0, \, j=1,\dots,N,
\end{equation}
be a bounded domain in $\R^n$, $n=2,3$, with
$C^2$-smooth boundary  $\partial\Omega=\cup_{j=0}^N\Gamma_j$ consisting of $N+1$
disjoint components $\Gamma_j=\partial\Omega_j$, $j=0,\dots, N$.
Consider  the stationary  Navier--Stokes system  with
nonhomogeneous boun-dary conditions
\begin{equation}
\label{NS}
 \left\{\begin{array}{rcl}
-\nu \Delta{\bf u}+\big({\bf u}\cdot \nabla\big){\bf u} +\nabla p &
= & {\bf f}\qquad \hbox{\rm in }\;\;\Omega,\\[4pt]
\div\,{\bf u} &  = & 0  \qquad \hbox{\rm in }\;\;\Omega,
\\[4pt]
 {\bf u} &  = & {\bf a}
 \qquad \hbox{\rm on }\;\;\partial\Omega.
\end{array}\right.
\end{equation}
The continuity
equation $(\ref{NS}_2)$ implies the compatibility
condition
\begin{equation}\label{flux}
\int\limits_{\partial\Omega}{\bf a}\cdot{\bf
n}\,ds=\sum\limits_{j=0}^N\int\limits_{\Gamma_j}{\bf a}\cdot{\bf
n}\,ds=\sum\limits_{j=0}^N\F_j=0
\end{equation}
necessary  for the solvability of problem (\ref{NS}), where ${\bf n}$ is a unit
 outward (with respect to $\Omega$) normal vector to
$\partial\Omega$ and $\F_j=\int_{\Gamma_j}{\bf a}\cdot{\bf
n}\,dS$.  Condition (\ref{flux}) means
that the total flux of the fluid through
$\partial\Omega$ is zero.

In his  famous  paper of 1933  \cite{Leray}   Jean Leray proved that problem (\ref{NS}) has a solution provided\footnote{Condition (\ref{flux0}) does not allow the presence of sinks and
sources. }
\begin{equation}\label{flux0}
{\F}_j=\int\limits_{\Gamma_j}{\bf a}\cdot{\bf n}\,dS=0,\qquad
j=0,1,\ldots,N.
\end{equation}
The case when the boundary value ${\bf a}$ satisfies only the necessary condition \eqref{flux} was left open by
Leray and the problem whether (\ref{NS}), \eqref{flux} admit (or do not admit) a solution  is  know in the scientific community as {\it Leray's problem}.

Leray's problem was studied in many
papers. However, in spite of all efforts, the existence of a weak
solution ${\bf u}\in W^{1,2}(\Omega)$  to problem (\ref{NS}) was
established  only under  assumption \eqref{flux0}
(see, e.g., \cite{Leray}, \cite{Lad1}, \cite{Lad}, \cite{VorJud}, \cite{KaPi1}), or
for sufficiently small fluxes ${\F}_j$ \footnote{This condition  does not
assumes the norm of the boundary value ${\bf a}$ to be small.}
(see, e.g.,  \cite{Finn}, \cite{Fu},
\cite{Galdi1}, \cite{Galdibook}, \cite{BOPI}, \cite{Russo}, \cite{RussoA}, \cite{Kozono}),   or under certain symmetry conditions on the domain $\Omega$ and the boundary value ${\bf a}$
(see, e.g.,   \cite{Amick}, \cite{Sazonov}, \cite{Fu1}, \cite{Morimoto}, \cite{Pukhnachev}, \cite{Pukhnachev1}).
Recently~\cite{kpr} the existence theorem for
(\ref{NS}) was  proved  for a plane domain $\Omega$ with two
connected components of the boundary  assuming only that the
flux through the external component is negative (inflow condition).
Similar  result was also obtained for the spatial axially symmetric
case \cite{kpr_a_arx}.
In particular, the existence was established without any restrictions on the fluxes $\F_j$, under the assumption that all components $\Gamma_j$ of
$\partial\Omega$ intersect the axis of symmetry.
For more detailed historical surveys  one can see the recent
papers~\cite{kpr} or \cite{Pukhnachev}--\cite{Pukhnachev1}.

In the present paper we solve Leray's problem for the  plane case
$n=2$ and for the axially symmetric  domains in $\R^3$.
 The main result for the plane case is as follows.

\begin{theo} \label{kmpTh4.1} {\sl Assume that  $\Omega\subset\R^2$
is a bounded domain of type \eqref{domain} with
$C^2$-smooth boundary $\partial\Omega$. If ${\bf f}\in W^{1,2}(\Omega)$  and  ${\bf a}\in
W^{3/2,2}(\partial\Omega)$  satisfies condition $(\ref{flux})$, then  problem $(\ref{NS})$ admits at least one
weak solution.}
\end{theo}

The proof of the existence theorem is based on  an a priori estimate
which we derive using a \textit{reductio ad absurdum} argument
of Leray \cite{Leray}. The essentially new part in
this argument is the use of Bernoulli's law obtained in~\cite{korob1} for Sobolev solutions to the Euler
equations (the detailed proofs are
presented in~\cite{kpr}). The results concerning Bernoulli's law are based
on the recent version of the~Morse-Sard theorem proved by J. Bourgain,
M.~Korobkov and J. Kristensen \cite{korob}. This theorem implies, in particular, that almost all level sets
of a function $\psi\in W^{2,1}(\Omega)$ are finite unions of $C^1$-curves. This allows  to construct suitable subdomains (bounded by smooth stream lines) and to estimate the $L^2$-norm of the gradient of the total head pressure. We use here some ideas which are close (on a~heuristic level) to the Hopf maximum principle for the solutions of elliptic PDEs (for a more detailed explanation see Subsection~\ref{subsub1}).
 Finally, a contradiction is obtained using the Coarea formula.

The paper is organized as follows. Section~2 contains
preliminaries. Basically, this section consists of standard facts,
except for the results of Subsection~2.2, where we formulate the
recent version~\cite{korob} of the Morse-Sard Theorem for the
space $W^{2,1}(\R^2)$, which plays a~key role. In
Subsection~3.1 we briefly recall the elegant \textit{reductio ad
absurdum} Leray's argument. In Subsection~3.2 we discuss properties of the
limit solution to the Euler equations, which were known before
(mainly, we recall some facts from~\cite{kpr}). In Subsection~3.3 we
prove some new properties of this limit solution and get a contradiction. Finally, in Section~4 we adapt these methods
to the~axially symmetric spatial case.

\section{Notation and auxiliary  results}
\subsection{Function spaces and definitions}
\setcounter{equation}{0}

By {\it a domain} we mean a connected open  set. Let
$\Omega\subset\R^n$, $n=2,3$, be a bounded domain with $C^2$-smooth
boundary~$\partial\Omega$. We use standard
notation for function spaces: $C^k(\overline\Omega)$,
$C^k(\partial\Omega)$, $W^{k,q}(\Omega)$, $\woq(\Omega)$,
$W^{\alpha,q}(\partial\Omega)$, where $\alpha\in(0,1),
k\in{\mathbb N}_0, q\in[1,+\infty]$.  In our notation we do not
distinguish function spaces for scalar and vector-valued
functions; it will be clear from the context whether we use scalar,
vector, or tensor-valued function spaces. Denote by $H(\Omega)$ the subspace
of all solenoidal  vector-fields ($\div{\bf u}=0$) from
$\wotwo(\Omega)$ equipped with the norm $\|{\bf
u}\|_{H(\Omega)}=\|\nabla{\bf u}\|_{L^2(\Omega)}$.  Observe that for
functions ${\bf u}\in H(\Omega)$ the norm $\|\cdot\|_{H(\Omega)}$
is equivalent to $\|\cdot\|_{W^{1,2}(\Omega)}$.

Working with Sobolev functions, we always assume that the "best
representatives" are chosen. For $w\in L^1_{\loc}(\Omega)$
the best representative $w^*$ is defined as
\begin{displaymath} w^*(x)=\left\{\begin{array}{rcl}\lim\limits_{r\to
0} \dashint_{B_r(x)}{
w}(z)dz, & {\rm \;if\; the\; finite\; limit\; exists;} \\[4pt]
 0 \qquad\qquad\quad & \; {\rm otherwise },
\end{array}\right.
\end{displaymath}
where
$\dashint_{B_r(x)}{ w}(z)dz=\frac{1}{\meas(
B_r(x))}\int_{B_r(x)}{ w}(z)dz$
and $B_r(x)=\{y: |y-x|<r\}$ is the ball
of radius $r$ centered at $x$.

Below we discuss some properties of the best representatives of
Sobolev functions.

\begin{lem}[see, for example, Theorem~1 of \S4.8 and Theorem~2 of \S{4.9.2} in \cite{evans}]
\label{kmpLem1} If $w\in W^{1,s}(\R^2)$,  $s\ge1$,   then there
exists a set $A_{1,w}\subset\R^2$ with the following properties:

{\rm (i)} $\mathfrak{H}^1(A_{1,w})=0$;

{\rm (ii)} for  each $x\in\Omega\setminus A_{1,w}$
\begin{displaymath}
\lim\limits_{r\to 0}\dashint\nolimits_{B_r(x)}|{
w}(z)-{w}(x)|^2\,dz=0;
\end{displaymath}

{\rm (iii)} for every $\varepsilon >0$ there exists a set $U\subset
\mathbb{R}^2$ with $\mathfrak{H}^1_\infty(U)<\varepsilon$ and
$A_{1,w}\subset U$ such that the function $w$ is continuous on
$\overline\Omega\setminus U$;

{\rm (iv)} for every  unit vector $\mathbf l\in \partial B_1(0)$  and almost all straight lines $L$ parallel to $\mathbf l$, the
restriction $w|_L$ is an absolutely continuous function $($of one
variable$)$.
\end{lem}

Here and henceforth we denote by $\mathfrak{H}^1$ the
one-dimensional Hausdorff measure, i.e.,
$\mathfrak{H}^1(F)=\lim\limits_{t\to 0+}\mathfrak{H}^1_t(F)$, where
$$\mathfrak{H}^1_t(F)=\inf\Big\{\sum\limits_{i=1}^\infty {\rm diam}
F_i:\, {\rm diam} F_i\leq t, F\subset \bigcup\limits_{i=1}^\infty
F_i\Big\}.$$

\begin{lemr} {\rm The property (iii) of Lemma~\ref{kmpLem1} means that
$f$ is quasicontinuous with respect
to the Hausdorff content $\mathfrak{H}^1_\infty$. Really, Theorem~1 (iii) of \S4.8
in \cite{evans} asserts that $f\in W^{1,s}(\R^2)$
is quasicontinuous with respect to the $s$-capacity.
But it is well known that for $s=1$ smallness of the
$1$-capacity of a set $F\subset\R^2$ is equivalent to
 smallness of $\mathfrak{H}^1_\infty(F)$
(see, e.g., Theorem~3 of \S5.6.3 in \cite{evans}
and its proof).}
\end{lemr}

\begin{lemr}
\label{kmpRem1} {\rm By the Sobolev extension theorem,  Lemma~\ref{kmpLem1} is true for functions $w\in
W^{1,s}(\Omega)$, where $\Omega\subset\R^2$ is a bounded Lipschitz  domain.
By the  trace theorem each function
$w\in W^{1,s}(\Omega)$ is "well-defined" for
$\mathfrak{H}^1$-almost all $x\in\partial\Omega$. Therefore,  we
assume that every function $w\in W^{1,s}(\Omega)$ is defined on
$\overline\Omega$.}
\end{lemr}

\subsection{On the Morse-Sard and Luzin N-properties of Sobolev functions in $W^{2,1}$}

First, let us  recall some classical differentiability properties of
Sobolev functions.

\begin{lem}[see Proposition~1 in \cite{Dor}]
\label{kmpThDor} {\sl If $\psi\in W^{2,1}(\R^2)$,  then $\psi$ is continuous and there
exists a set $A_{\psi}$ with $\mathfrak{H}^1(A_{\psi})=0$ such that
$\psi$ is differentiable (in the classical sense) at all $x\in\R^2\setminus A_{\psi}$.
Moreover, the classical derivative coincides with
$\nabla\psi(x)$, where $\lim\limits_{r\to 0}\dashint\nolimits_{B_r(x)}|
\nabla\psi(z)-\nabla\psi(x)|^2\,dz=0$.}
\end{lem}

The theorem below is due to J. Bourgain,
M.~Korobkov and J. Kristensen \cite{korob}.

\begin{theo}
\label{kmpTh1.1} {\sl Let  $\Omega\subset\R^2$ be a bounded domain with
Lipschitz boundary. If $\psi\in W^{2,1}(\Omega)$, then

{\rm (i)}
$\mathfrak{H}^1(\{\psi(x)\,:\,x\in\overline{\Omega}\setminus
A_\psi\,\,\&\,\,\nabla \psi(x)=0\})=0$;

{\rm (ii)} for every $\varepsilon>0$ there exists $\delta>0$ such
that $\mathfrak{H}^1(\psi(U))<\varepsilon$ for any set $U\subset \overline{\Omega}$ with
$\mathfrak{H}^1_\infty(U)<\delta$;

{\rm (iii)} for every $\varepsilon>0$ there exists an open set
$V\subset\mathbb{R}$ with $\mathfrak{H}^1(V)<\varepsilon$ and a function $g\in C^1(\mathbb{R}^2)$ such
that for each
$x\in\overline\Omega$ if $\psi(x)\notin V$, then $x\notin
A_{\psi}$
and $\psi(x)=g(x)$, $\nabla \psi(x)=\nabla g(x)\neq 0$;

{\rm (iv)}  for $\mathfrak{H}^1$--almost all $y\in
\psi(\overline{\Omega})\subset \mathbb{R}$ the preimage
$\psi^{-1}(y)$ is a finite disjoint family of $C^1$-curves $S_j$,
$ j=1, 2, \ldots, N(y)$. Each $S_j$ is either a cycle in
${\Omega}$ $($i.e., $S_j\subset{\Omega}$ is homeomorphic to the
unit circle $\mathbb{S}^1)$ or  a simple arc with endpoints
on $\partial{\Omega}$ $($in this case $S_j$ is transversal to
$\partial{\Omega}\,)$.}
\end{theo}

\subsection{Some facts from topology}
\label{Kronrod-s}

We shall need  some topological
definitions and results. By  {\it continuum} we mean a compact
connected set. We understand  connectedness  in the sense of
general topology. A set is called {\it an arc} if it is
homeomorphic to the unit interval~$[0,1]$.

Let us shortly present  some results from the classical paper of A.S.~Kron-\\rod~
\cite{Kronrod} concerning level sets of continuous functions. Let
${Q}=[0,1]\times[0,1]$ be a square in $\mathbb{R}^2$ and let $f$ be a
continuous function  on ${Q}$. Denote by $E_t$ a level set
of the function $f$, i.e., $E_t=\{x\in{Q}: f(x)=t\}$. A component
$K$  of the level set $E_t$ containing a point $x_0$ is a maximal
connected subset of $E_t$ containing $x_0$. By $T_f$ denote a
family of all connected components of level sets of~$f$. It was established in~\cite{Kronrod} that $T_f$ equipped by a
natural topology is a tree. Vertices of this tree are the components~$C\in T_f$ which do not separate~$Q$, i.e.,
$Q\setminus C$ is a connected set. Branching points of the tree are the components $C\in T_f$ such that $Q\setminus C$ has more than two connected components. By results of \cite{Kronrod}, see also
\cite{Moore} and \cite{Pittman}, the set of all branching points of~$T_f$
is at most countable. The main property of a tree is that any two points could be joined by a unique arc.
Therefore, the same is true for~$T_f$.

\begin{lem} [\cite{Kronrod}]
\label{kmpLem6}
If $f\in C(Q)$, then
for any two different points  $A\in T_f$ and $B\in T_f$, there exists a unique arc
$J=J(A,B)\subset T_f$ joining $A$ to $B$. Moreover, for every inner
point $C$ of this arc the points $A,B$ lie in  different
connected components of the set $T_f\setminus\{C\}$.
\end{lem}

We can reformulate the above Lemma in the following equivalent
form.

\begin{lem} \label{kmpLem7}{\sl If  $f\in C(Q)$, then for
any two different points $A,B\in T_f$, there exists an injective function
$\varphi:[0,1]\to T_f$ with the properties

{\rm (i)} $\varphi(0)=A$, $\varphi(1)= B$;

{\rm (ii)} for any $t_0\in[0,1]$, 
$$
\lim\limits_{[0,1]\ni
t\to t_0}\sup\limits_{x\in \varphi(t)}\dist(x,\varphi(t_0))\to0;
$$

{\rm (iii)}  for any $t\in(0,1)$ the sets $A,B$ lie in different
connected components of the set \ $Q\setminus\varphi(t)$.}
\end{lem}

\begin{lemr}
\label{kmpRem2} {\rm
If in  Lemma~\ref{kmpLem7} $f\in W^{2,1}(Q)$, then by
Theorem~\ref{kmpTh1.1}~(iv), there exists a dense subset $E$ of
$(0,1)$ such that $\varphi(t)$ is a $C^1$-- curve for every $t\in E$.
Moreover, $\varphi(t)$ is either a cycle   or a simple
arc with endpoints on $\partial Q$.}
\end{lemr}

\begin{lemr}
\label{kmpRem1.2} {\rm All results of
Lemmas~\ref{kmpLem6}--\ref{kmpLem7} remain valid for level sets of
continuous functions $f:\overline\Omega\to\R$, where $\Omega$ is a
multi--connected bounded domain of type~(\ref{domain}), provided
$f\equiv\xi_j=\const$ on each inner boundary component $\Gamma_j$ with $j=1,\dots,N$.
Indeed, we can extend~$f$ to the whole $\overline\Omega_0$ by putting $f(x)=\xi_j$ for $x\in \overline\Omega_j$, $j=1,\dots,N$.
The extended function~$f$ will be continuous on the~set $\overline\Omega_0$ which is homeomorphic to the unit square
$Q=[0,1]^2$.
}
\end{lemr}
\

\section{The plane case}
\label{ppPlane} \setcounter{theo}{0} \setcounter{lem}{0}
\setcounter{lemr}{0}\setcounter{equation}{0}

\subsection{Leray's argument ``reductio ad absurdum'' }
\label{Leray_arg}

Consider the Navier--Stokes problem (\ref{NS}) in the $C^2$-smooth  domain
$\Omega\subset\R^2$  defined by (\ref{domain}) with ${\bf f}\in W^{1,2}(\Omega)$.  Without loss of generality, we may assume that ${\bf f}=\nabla^\perp b$  with $b\in W^{2,2}(\Omega)$\footnote{By the Helmholtz-Weyl decomposition, for a $C^2$-smooth bounded domain $\Omega\subset\R^n$, $n=2,3$, every
$\fe\in W^{1,2}(\Omega)$ can be represented as the sum $\fe={\rm curl}\,{\bf b}+\nabla \varphi$ for $n=3$, and $\fe=\nabla^\perp b+\nabla \varphi$ with ${\bf b}, b,\,\varphi\in W^{2,2}(\Omega)$, and
the gradient part is included then into the pressure term (see, e.g., \cite{Lad}).}, where
$(x,y)^\bot=(-y,x)$.  If the boundary value ${\bf
a}\in W^{3/2,2}(\partial\Omega)$ satisfies condition (\ref{flux}),
then there exists a solenoidal extension ${\bf A}\in
W^{2,2}(\Omega)$ of ${\bf a}$ (see \cite{Lad}, \cite{Temam}, \cite{Galdibook}). Using this fact and standard results \cite{Lad}, we can find a weak solution ${\bf U}\in
W^{2,2}(\Omega)$ to the Stokes problem such that ${\bf U}-{\bf A}\in
H(\Omega)\cap W^{2,2}(\Omega)$ and
\begin{equation}\label{4.1}
\nu\int\limits_\Omega\nabla{\bf U}\cdot\nabla\bfeta\,dx= \int\limits_\Omega{\bf f}\cdot\bfeta\,dx
\quad\forall\;\bfeta\in H(\Omega).
\end{equation}
Moreover,
\begin{equation}\label{4.2}
\|{\bf U}\|_{W^{2,2}(\Omega)}\leq c\big(\|{\bf
a}\|_{W^{3/2,2}(\partial\Omega)}+\|{\bf f}\|_{L^2(\Omega)}\big).
\end{equation}

By {\it weak solution} of problem (\ref{NS}) we understand a
function ${\bf u}$ such that ${\bf w}={\bf u}-{\bf U}\in
H(\Omega)$ and
\begin{displaymath}
\nu\int\limits_\Omega\nabla{\bf
w}\cdot\nabla\bfeta\,dx-\int\limits_\Omega\big(({\bf w}+{\bf
U})\cdot\nabla\big)\bfeta\cdot{\bf
w}\,dx-\int\limits_\Omega\big({\bf
w}\cdot\nabla\big)\bfeta\cdot{\bf U}\,dx
\end{displaymath}
\begin{equation}\label{4.3}
=\int\limits_\Omega\big({\bf U}\cdot\nabla\big)\bfeta\cdot{\bf
U}\,dx \qquad\forall\bfeta\in H(\Omega).
\end{equation}

Let us reproduce shortly the contradiction
argument of Leray \cite{Leray} which was later used
in many other papers (see, e.g., \cite{Lad1}, \cite{Lad},
\cite{KaPi1}, \cite{Amick}; see also \cite{kpr} for details).
It is well known (see, e.g., \cite{Lad}) that
integral identity (\ref{4.3}) is equivalent to an operator
equation in the space $H(\Omega)$ with a compact operator.
Therefore, by the Leray--Schauder theorem, to prove the 
existence of a weak solution to Navier--Stokes problem (\ref{NS}),
it is sufficient to show that all the solutions of the
integral identity
\begin{displaymath}
\nu\int\limits_\Omega\nabla{\bf
w}\cdot\nabla\bfeta\,dx-\lambda\int\limits_\Omega\big(({\bf
w}+{\bf U})\cdot\nabla\big)\bfeta\cdot{\bf
w}\,dx-\lambda\int\limits_\Omega\big({\bf
w}\cdot\nabla\big)\bfeta\cdot{\bf U}\,dx
\end{displaymath}
\begin{equation}\label{4.4}
=\lambda\int\limits_\Omega\big({\bf
U}\cdot\nabla\big)\bfeta\cdot{\bf U}\,dx \qquad \forall\;\bfeta\in
H(\Omega)
\end{equation}
are uniformly  bounded in
$H(\Omega)$ (with respect to $\lambda\in[0,1]$). Assume that this is false. Then there exist sequences $
\{\lambda_k\}_{k\in{\mathbb{N}}}\subset [0, 1]$ and $\{\widehat{\bf
w}_k\}_{k\in\mathbb{N}}\in H(\Omega)$ such that
\begin{displaymath}
\nu\int\limits_\Omega\nabla\widehat{\bf
w}_k\cdot\nabla\bfeta\,dx-\lambda_k\int\limits_\Omega\big((\widehat{\bf
w}_k+{\bf U})\cdot\nabla\big)\bfeta\cdot\widehat{\bf
w}_k\,dx-\lambda_k\int\limits_\Omega\big(\widehat{\bf
w}_k\cdot\nabla\big)\bfeta\cdot{\bf U}\,dx
\end{displaymath}
\begin{equation}\label{4.5}
=\lambda_k\int\limits_\Omega\big({\bf
U}\cdot\nabla\big)\bfeta\cdot{\bf U}\,dx\qquad\forall\,\bfeta\in
H(\Omega),
\end{equation}
and
\begin{equation}\label{4.6}
\lim\limits_{k\to\infty}\lambda_k=\lambda_0\in[0, 1],\quad
\lim\limits_{k\to\infty}J_k=\lim\limits_{k\to\infty}\|\widehat{\bf
w}_k\|_{H(\Omega)}=\infty.
\end{equation}

Using well known techniques (\cite{kpr}, \cite{Amick}),
one shows that there exist  $\widehat p_k$ with\footnote{The uniform estimates for the norms
$\|p_k\|_{W^{1,q}(\Omega)}$ follow from
well-known results concerning  regularity of solutions to the Stokes
problem (see \cite[Chapter 1, \S2.5]{Temam} or \cite{Lad}). Observe that in
\cite{kpr} we could have only $p_k\in W^{1,q}_\loc(\Omega)$ because
$\partial\Omega$ has been assumed to be only Lipschitz. However, for
domains $\Omega$ with $C^2$-smooth boundary and ${\bf a}\in
W^{3/2,2}(\partial\Omega)$ the corresponding estimates hold globally.} $\|\widehat p_k\|_{W^{1,q}(\Omega)}\le C(q)J_k^2$, $q\in[1,2)$, such that the pair $\big(\widehat\ue_k= \widehat{\bf
w}_k+{\bf U}, \ \widehat p_k\big)$ is a solution to the following system

\begin{equation}\label{NSk-im}
\left\{\begin{array}{rcl}-\nu\Delta\widehat{\bf u}_k +\lambda_k\big(\widehat{\bf
u}_k\cdot\nabla\big)\widehat{\bf u}_k+\nabla \widehat p_k & = & \fe\qquad
\hbox{\rm in }\;\;\Omega,
\\[4pt]
\div\widehat{\bf u}_k & = & 0 \;\qquad \hbox{\rm in }\;\;\Omega,
\\[4pt]
\widehat{\bf u}_k &  = & {\bf a}
 \quad\  \hbox{\rm on }\;\;\partial\Omega.
\end{array}\right.
\end{equation}

Choose $\bfeta=J_k^{-2}\widehat{\bf w}_k$ in (\ref{4.5})  and set
${\bf w}_k=J_k^{-1}\widehat{\bf w}_k$. Taking into account that
\begin{displaymath}
\int\limits_\Omega\big(({\bf w}_k+{\bf U})\cdot\nabla\big){\bf
w}_k\cdot{\bf w}_k\,dx=0,
\end{displaymath}
we have
\begin{equation}\label{4.7}
\nu\int\limits_\Omega|\nabla{\bf
w}_k|^2\,dx=\lambda_k\int\limits_\Omega\big({\bf
w}_k\cdot\nabla\big){\bf w}_k\cdot{\bf U}\,dx+
J_k^{-1}\lambda_k\int\limits_\Omega\big({\bf
U}\cdot\nabla\big){\bf w}_k\cdot{\bf U}\,dx.
\end{equation}
Since $\|{\bf w}_k\|_{H(\Omega)}=1$, there exists a
subsequence $\{{\bf w}_{k_l}\}$   converging weakly in
$H(\Omega)$ to a vector field ${\bf v}\in H(\Omega)$. By
the compact embedding
\begin{displaymath}
H(\Omega)\hookrightarrow L^r(\Omega) \quad
\forall\,r\in[1,\infty),
\end{displaymath}
the subsequence $\{{\bf w}_{k_l}\}$ converges strongly in
$L^r (\Omega)$. Therefore, letting $k_l\to\infty$ in
equality (\ref{4.7}), we obtain
\begin{equation}\label{4.8}
\nu=\lambda_0\int\limits_\Omega\big({\bf v}\cdot\nabla\big){\bf
v}\cdot{\bf U}\,dx.
\end{equation}
In particular, $\lambda_0>0$, so $\lambda_k$ are separated from
zero.

Put $\nu_k=(\lambda_kJ_k)^{-1}\nu$. Multiplying identities~(\ref{NSk-im}) by
$\frac{1}{\lambda_kJ^2_k}=\frac{\lambda_k\nu^2_k}{\nu^2}$, we see that the pair $\big(\ue_k= \frac1{J_k}\widehat{\ue}_k, \ p_k=\frac{1}{\lambda_kJ^2_k}\widehat p_k\big)$ satisfies the following system

\begin{equation}\label{NSk}
\left\{\begin{array}{rcl}-\nu_k\Delta{\bf u}_k +\big({\bf
u}_k\cdot\nabla\big){\bf u}_k+\nabla p_k & = & \fe_k\qquad
\hbox{\rm in }\;\;\Omega,
\\[4pt]
\div{\bf u}_k & = & 0 \;\qquad \hbox{\rm in }\;\;\Omega,
\\[4pt]
{\bf u}_k &  = & {\bf a}_k
 \quad\  \hbox{\rm on }\;\;\partial\Omega,
\end{array}\right.
\end{equation}
where $\fe_k=\frac{\lambda_k\nu_k^2}{\nu^2}\,{\bf f}$, \ ${\bf a}_k=\frac{\lambda_k\nu_k}\nu\,{\bf a}$, the norms
$\|\ue_k\|_{W^{1,2}(\Omega)}$ and  $\|p_k\|_{W^{1,q}(\Omega)}$ are
uniformly bounded for each $q\in[1,2)$, $\ue_k\in W^{3,2}_{\loc}(\Omega)$,
$p_k\in W^{2,2}_{\loc}(\Omega)$\footnote{The interior  regularity of the solution depends on the regularity of $\fe\in W^{1,2}(\Omega)$,
but not on the regularity of the boundary value  ${\bf a}$, \ see \cite{Lad}.}, and  $\ue_k\rightharpoonup \ve\mbox{ \ in \
}W^{1,2}(\Omega),\quad p_k\rightharpoonup p\mbox{ \ in \
}W^{1,q}(\Omega)$.
Moreover, the limit functions $(\ve,p)$ satisfy the Euler system
\begin{equation}
\label{2.1}\left\{\begin{array}{rcl} \big({\bf
v}\cdot\nabla\big){\bf v}+\nabla p & = & 0 \qquad \ \ \
\hbox{\rm in }\;\;\Omega,\\[4pt]
\div{\bf v} & = & 0\qquad \ \ \ \hbox{\rm in }\;\;\Omega,
\\[4pt]
{\bf v} &  = & 0\ \
 \qquad\  \hbox{\rm on }\;\;\partial\Omega.
\end{array}\right.
\end{equation}

In conclusion,  we can state the following lemma.

\begin{lem}
\label{lem_Leray} {\sl Assume that  $\Omega\subset\R^2$ is a bounded domain of type \eqref{domain} with
$C^2$-smooth boundary $\partial\Omega$,  $\fe=\nabla^\bot b$, $b\in W^{2,2}(\Omega)$, and  ${\bf a}\in W^{3/2,2}(\partial\Omega)$ satisfies condition \eqref{flux}.  If there are no weak solutions to~\eqref{NS}, then there exist $ \ve, p$ with the following properties.

\medskip

(E)  \ \,  $\ve\in
W^{1,2}(\Omega)$, $p\in W^{1,q}(\Omega)$, $q\in(1,2)$, and the pair $\big(\ve, p\big)$ satisfies the~Euler system (\ref{2.1}).

\medskip

(E-NS) \ \,Conditions (E) are satisfied and there exist
sequences of functions $\ue_k \in {W^{1,2}(\Omega)}$, $p_k\in
{W^{1,q}(\Omega)}$ and numbers $\nu_k\to 0+$,
$\lambda_k\to\lambda_0>0$ such that the norms
$\|\ue_k\|_{W^{1,2}(\Omega)}$, $\|p_k\|_{W^{1,q}(\Omega)}$ are
uniformly bounded for every  $q\in[1,2)$,  the pairs $(\ue_k,p_k)$ satisfy
(\ref{NSk}) with $\fe_k=\frac{\lambda_k\nu_k^2}{\nu^2}\,{\bf f}$, \ ${\bf a}_k=\frac{\lambda_k\nu_k}\nu\,{\bf a}$, and
$$
\|\nabla \ue_k\|_{L^2(\Omega)}\to1,\quad\ue_k\rightharpoonup \ve\mbox{ \ in \
}W^{1,2}(\Omega),\quad p_k\rightharpoonup p\mbox{ \ in \
}W^{1,q}(\Omega)\quad\forall \;q\in[1,2).
$$
Moreover, $\ue_k\in W^{3,2}_{\loc}(\Omega)$, $p_k\in W^{2,2}_{\loc}(\Omega)$.}
\end{lem}

From now on we assume that  assumptions (E-NS)
are satisfied.  Our goal is to prove that they   lead to a
contradiction. This implies the validity of
Theorem~\ref{kmpTh4.1}.

\subsection{Some previous results on the Euler equations}
\label{EPprev}

In this subsection  we collect the information on the~limit solution $\big(\ve, p\big)$ to \eqref{2.1} obtained in previous papers. The next statement was proved in
\cite[Lemma 4]{KaPi1} and in \cite[Theorem 2.2]{Amick} (see also
\cite[Remark 3.2]{kpr}).

\begin{theo}
\label{kmpTh2.3'} {\sl If conditions {\rm (E)} are satisfied, then  there exist constants $\widehat p_0,\dots,\widehat p_N$
such that
\begin{equation} \label{bp2}
p(x)\equiv \widehat p_j\quad\mbox{for }\Ha^1-\mbox{almost all }
x\in\Gamma_j.\end{equation}}
\end{theo}

\begin{lemA}\label{ppp1} {\sl If conditions {\rm (E-NS)} are
satisfied, then
\begin{equation}\label{4.8'''}
-\frac{\nu}{\lambda_0}=\sum\limits_{j=0}^N\widehat p_j
\int\limits_{\Gamma_j}{\bf a}\cdot{\bf
n}\,ds=\sum\limits_{j=0}^N\widehat p_j \mathcal F_j.
\end{equation}
}
\end{lemA}

\pr By simple calculations from (\ref{4.8}) and (\ref{2.1}${}_1$) it follows
%\begin{equation}\label{4.8''}
$$
\frac{\nu}{\lambda_0}=-\int\limits_\Omega\nabla p\cdot{\bf
U}\,dx=- \int\limits_\Omega\div(p{\bf
U})\,dx=-\int\limits_{\partial\Omega}p\,{\bf a}\cdot{\bf n}\,ds.
%\end{equation}
$$
In virtue of~(\ref{bp2}), this implies~(\ref{4.8'''}).  $\qed$

\medskip

Set $\Phi_k=p_k+\frac{1}2|\ue_k|^2$, \
$\Phi=p+\frac{1}2|\ve|^2$. From (\ref{2.1}${}_2$) and
(\ref{2.1}${}_3$) it follows that there exists a stream function
$\psi\in W^{2,2}(\Omega)$ such that
\begin{equation}\label{STR_grad}
\nabla\psi\equiv\ve^\bot\quad\mbox{ in }\overline\Omega.
\end{equation}
Here and henceforth we set $(a,b)^\bot=(-b,a)$.

Applying Lemmas~\ref{kmpLem1}, \ref{kmpThDor} and
Remark~\ref{kmpRem1} to the functions $\ve,\psi,\Phi$ we get the
following

\begin{lem}
\label{kmpTh2.1} {\sl If conditions {\rm (E)} are satisfied, then the stream function~$\psi$ is continuous
on~$\overline\Omega$ and
there exists a set $A_{\ve}\subset \overline\Omega$ such that

 {\rm (i)}\quad $ \mathfrak{H}^1(A_{\ve})=0$;

{\rm (ii)} for  all  $x\in\Omega\setminus A_{\ve}$
\begin{displaymath}
\lim\limits_{r\to
0}\dashint\nolimits_{B_r(x)}|\ve(z)-\ve(x)|^2dz=\lim\limits_{r\to
0}\dashint\nolimits_{B_r(x)}|{\Phi}(z)-{\Phi}(x)|^2dz=0;
\end{displaymath}
moreover, the function $\psi$ is differentiable at $x$ and
$\nabla\psi(x)=(-v_2(x), v_1(x))$;

{\rm  (iii) } for every  $\varepsilon >0$ there exists a set
$U\subset \mathbb{R}^2$ with
$\mathfrak{H}^1_\infty(U)<\varepsilon$ such that $A_{\ve}\subset U$ and the
functions $\ve, \Phi$ are continuous in $\overline\Omega\setminus
U$.}
\end{lem}
\vspace{0.3cm}

The next version of Bernoulli's Law for solutions in Sobolev
spaces was obtained in~\cite[Theorem~1]{korob1} (see also
\cite[Theorem 3.2]{kpr} for a more detailed proof).

\begin{theo}
\label{kmpTh2.2} {\sl Let conditions {\rm (E)} be satisfied and let
 $A_\ve\subset \overline\Omega$ be the  set 
from   Lemma~\ref{kmpTh2.1}. For any compact
connected  set $K\subset \overline\Omega$ the following
property holds: if
\begin{equation}
\label{2.4} \psi\big|_{K}=\const,
\end{equation}
then
\begin{equation}
\label{2.5'}  \Phi(x_1)=\Phi(x_2) \quad\mbox{for
 all \,}x_1,x_2\in K\setminus A_{\bf v}.
\end{equation}
}
\end{theo}

\begin{lem}
\label{lkrSTR} {\sl If conditions {\rm (E)} are satisfied, then
there exist constants $\xi_0,\dots, \xi_N\in\R$ such that
$\psi(x)\equiv \xi_j$ on each component $\Gamma_j$,
$j=0,\dots,N$.}
\end{lem}

\pr Consider any boundary component~$\Gamma_j$. Since $\psi$ is continuous on~$\overline\Omega$ and ~$\Gamma_j$ is connected, we have that $\psi(\Gamma_j)$ is also a connected set. On the other hand,
since $\nabla\psi(x)=0$ for $\Ha^1$-almost all $x\in\Gamma_j$ (see~(\ref{2.1}${}_3$) and (\ref{STR_grad})\,), Theorem~\ref{kmpTh1.1} (i)--(ii)
yields $\Ha^1(\psi(\Gamma_j))=0$. Therefore, $\psi(\Gamma_j)$ is a singleton.
$\qed$

\medskip

For $x\in\overline\Omega$ denote by  $K_x$ the connected component
of the level set $\{z\in\overline\Omega:\psi(z)=\psi(x)\}$
containing the point $x$.  By Lemma~\ref{lkrSTR}, $K_x\cap\partial\Omega=\emptyset$ for every
$y\in\psi(\overline\Omega)\setminus\{\xi_0,\dots,\xi_N\}$ and
for every~ $x\in \psi^{-1}(y)$. Thus,
Theorem~\ref{kmpTh1.1}~(ii),~(iv) implies that for almost all $y\in\psi(\overline\Omega)$ and
for every~$x\in \psi^{-1}(y)$ the equality $K_x\cap A_{\ve}=\emptyset$ holds
and the component $K_x\subset\Omega$ is a $C^1$-- curve homeomorphic to the
circle. We call such  $K_x$ {\it an admissible cycle}.

The
next lemma was obtained in~\cite[Lemma 3.3]{kpr}.

\begin{lem}
\label{kmpLem2.11} {\sl If conditions {\rm (E-NS)} are satisfied,
then the sequence $\{\Phi_k|_S\}$ converges
to $\Phi|_S$ uniformly $\Phi_k|_S\rightrightarrows\Phi|_S$ on almost
all\,\footnote{``Almost all cycles'' means cycles in preimages
$\psi^{-1}(y)$ for almost all values $y\in\psi(\overline\Omega)$.}
admissible cycles $S$.}
\end{lem}

 Admissible cycles $S$ from  Lemma~\ref{kmpLem2.11} will be called  {\it regular cycles}.

\subsection{Obtaining a contradiction}
\label{EPcontr}

We consider two cases.

(a) The maximum of $\Phi$ is attained on the boundary $\partial\Omega$:
\begin{equation}\label{as-prev0-0} \max\limits_{j=0,\dots,N}\widehat
p_j=\esssup\limits_{x\in\Omega}\Phi(x).
\end{equation}

(b) The  maximum of $\Phi$ is not attained\footnote{The case $\esssup\limits_{x\in\Omega}\Phi(x)=+\infty$ is not excluded. } on $\partial\Omega$:
\begin{equation}\label{as-prev-id} \max\limits_{j=0,\dots,N}\widehat
p_j<\esssup\limits_{x\in\Omega}\Phi(x).
\end{equation}

\subsubsection{The maximum of $\Phi$ is attained on the boundary $\partial\Omega$}\label{subsub1}
Let  \eqref{as-prev0-0} hold.
Adding a constant to the pressure we can assume, without loss of generality, that
\begin{equation}\label{as-prev0} \max\limits_{j=0,\dots,N}\widehat
p_j=\esssup\limits_{x\in\Omega}\Phi(x)=0.
\end{equation}
In particular,
\begin{equation}\label{as1'}
\Phi(x)\le0\quad\mbox{in }\Omega.
\end{equation}

If  $\widehat p_0=\widehat p_1=\dots=\widehat p_N$, then by Corollary~\ref{ppp1} and the flux condition \eqref{flux},
we immediately obtain the required contradiction.
Thus, assume that
\begin{equation}\label{as-prev-1}
\min\limits_{j=0,\dots,N}\widehat p_j<0.
\end{equation}

Change (if necessary) the numbering of the boundary components
$\Gamma_0$, $\Gamma_1$, \dots, $\Gamma_N$ in such a way  that
\begin{equation}\label{as01}
\widehat p_j<0, \quad j=0,\dots,M,
\end{equation}
\begin{equation}\label{as1}
\widehat p_{M+1}=\dots=\widehat p_N=0.
\end{equation}

First, we introduce the main idea of the proof in a
heuristic way. It is well known that every $\Phi_k$ satisfies the
linear elliptic equation
\begin{equation}\label{cle_laps}
\Delta\Phi_k=\omega_k^2+\frac1{\nu_k}\div(\Phi_k\ue_k)-\frac{1}{\nu_k}{\bf f}_k\cdot\ue_k
\end{equation}
If $\fe_k=0$,  then by
Hopf's maximum principle, in  a~subdomain
$\Omega'\Subset\Omega$ with $C^2$-- smooth boundary~$\partial\Omega'$ the maximum of $\Phi_k$ is attained
at the boundary $\partial\Omega'$, and if $x_*\in\partial\Omega'$ is a maximum point, then  the normal
derivative of $\Phi_k$ at $x_*$ is strictly positive. It is not
sufficient to apply this property directly. Instead  we will use  some
''integral analogs'' that lead to a contradiction by using the the Coarea
formula (see Lemmas~\ref{lkr11}--\ref{lem_Leray_fc}).
For $i\in\N$ and sufficiently large $k\ge k(i)$ we construct a~set $E_i\subset \Omega$
consisting of level lines of $\Phi_k$ such that $\Phi_k|_{E_i}\to0$ as $i\to\infty$ and $E_i$
separates the boundary component $\Gamma_N$ (where $\Phi=0$) from
the boundary components $\Gamma_j$ with $j=0,\dots,M$ (where $\Phi<0$). On the one hand, the
length of each of these level lines is bounded from below by a positive constant (since they separate
the boundary components), and by the Coarea formula this implies  the estimate from below
for $\int_{E_i}|\nabla \Phi_k|$. On the other hand, elliptic equation (\ref{cle_laps})
for $\Phi_k$, the convergence $\fe_k\to 0$, and  boundary conditions~(\ref{NSk}${}_3$) allow us
 to estimate $\int_{E_i}|\nabla \Phi_k|^2$ from above (see Lemma~\ref{lkr11}), and this   asymptotically contradicts
the previous one.

The main idea of the proof for a general multiply connected domain  is the same as in the case of annulus--like domains (when $\partial\Omega=\Gamma_0\cup\Gamma_1$\,).
The proof has an analytical nature and unessential differences concern only well known geometrical properties of
level sets of continuous functions of two variables.

First of all, we need some information concerning the behavior of the limit total head pressure~$\Phi$
on stream lines. We do not know whether the function~$\Phi$ is continuous or not on $\Omega$.
But we shall prove that~$\Phi$ has some continuity properties on stream lines.

By Remark~\ref{kmpRem1.2} and Lemma~\ref{lkrSTR}, we can apply
Kronrod's results to the stream function~$\psi$. Define the
total head pressure on the Kronrod tree $T_\psi$ (see
Subsection~\ref{Kronrod-s} ) as follows. Let $K\in T_\psi$ with $\diam
K>0$. Take any $x\in K\setminus A_\ve$ and put $\Phi(K)=\Phi(x)$.
This definition is correct by Bernoulli's Law (see
Theorem~\ref{kmpTh2.2}).
\begin{lem}
\label{lkr1} Let $A,B\in T_\psi$, $\diam A>0, \diam B>0$. Consider
the corresponding arc $[A,B]\subset T_\psi$ joining $A$ to $B$
$($see Lemmas~$\ref{kmpLem6}-\ref{kmpLem7}$ $)$. Then the
restriction $\Phi|_{[A,B]}$ is a continuous function.
\end{lem}

\pr Put $(A,B)=[A,B]\setminus\{A,B\}$.
Let $C_i\in(A,B)$ and $C_i\to C_0$ in
$T_\psi$. By construction, each $C_i$ is a connected component of
the level set of $\psi$ and the sets $A,B$ lie in different
connected components of $\R^2\setminus C_i$. Therefore,
\begin{equation} \label{KRT0}
\diam(C_i)\ge\min(\diam(A),\diam(B))>0.\end{equation} By the
definition of convergence in $T_\psi$, we have
\begin{equation}
\label{KRT1} \sup\limits_{x\in C_i}\dist(x, C_0)\to 0\quad\mbox{
as } i\to\infty.
\end{equation}
 By Theorem~\ref{kmpTh2.2}, there
exist constants $c_i\in\R$ such that $\Phi(x)\equiv c_i$ for all
$x\in C_i\setminus A_\ve$, where $\Ha^1(A_\ve)=0$. Analogously,
$\Phi(x)\equiv c_0$ for all $x\in C_0\setminus A_\ve$. If $c_i\nrightarrow c_0$, then
we can assume, without loss of generality, that
\begin{equation}
\label{cont2} c_i\to c_\infty\ne c_0\quad\mbox{as }i\to\infty
\end{equation}
and the components $C_i$  converge as $i\to\infty$ in the Hausdorff metric\footnote{The Hausdorff
distance $d_H$ between two compact sets $A,B\subset\R^n$ is
defined as follows: $d_H(A,B) = \max\bigl(\sup\limits_{a\in A}
\dist(a,B), \sup\limits_{b\in B} \dist(b,A)\bigr)$ (see, e.g., \S
7.3.1 in \cite{metric}). By Blaschke selection theorem [ibid], for any uniformly bounded
sequence of compact sets $A_i\subset\R^n$ there exists a
subsequence $A_{i_j}$ which converges to some compact set $A_0$
with respect to the Hausdorff distance. Of course, if all $A_i$
are compact connected sets and $\diam A_i\ge\delta$ for some
$\delta>0$, then the limit set $A_0$ is also connected and $\diam
A_0\ge\delta$.}
 to some
set~$C'_0\subset C_0$. Clearly, $\diam(C'_0)>0$.
Take a straight line
$L$
such that the projection of $C'_0$ on $L$ is not a singleton. Since
$C'_0$ is a connected set,  this projection is a
segment. Let  $I_0$ be the interior of this segment. For $z\in
I_0$ by $L_z$ denote the straight line such that $z\in L_z$ and
$L_z\perp L$. From Lemma~\ref{kmpTh2.1}~(i),\,(iii) it follows that $L_z\cap A_\ve=\emptyset$  for
$\mathfrak H^1$-almost all $z\in I_0$,
and the restriction $\Phi|_{\overline\Omega\cap L_z}$ is
continuous. Fix a point $z\in I_0$ with above properties.  Then by
construction $C_i\cap
L_z\ne\emptyset$ for sufficiently
large~$i$. Now, take a sequence $y_i\in C_i\cap L_z$ and extract a convergent subsequence
$y_{i_j}\to y_0\in C'_0$.
Since $\Phi|_{\overline\Omega\cap L_z}$ is continuous, we have
$\Phi(y_{i_j})=c_{i_j}\to\Phi(y_0)=c_0$ as $j\to\infty$. This contradicts~(\ref{cont2}). $\quad\qed$

\medskip

For the velocities $\ue_k=(u_k^1,u_k^2)$ and $\ve=(v^1,v^2)$ denote by
$\omega_k$ and $\omega$ the corresponding vorticities:
$\omega_k=\partial_2u^1_k-\partial_1u^2_k$, \
$\omega=\partial_2v^1-\partial_1v^2=\Delta\psi$. The following
formulas are direct consequences of $(\ref{2.1})$, $(\ref{NSk})$:
\begin{equation}\label{eu0}
\nabla\Phi\equiv\omega\ve^\bot=\omega\nabla\psi,\qquad
\nabla\Phi_k\equiv-\nu_k\nabla^\bot\omega_k+\omega_k\ue^\bot_k+\fe_k
\quad\mbox{ in }\Omega.\end{equation}

We say that a set $\mathcal Z\subset T_\psi$ has $T$-measure
zero if $\Ha^1(\{\psi(C):C\in \mathcal Z\})=0$. The function $\Phi|_{T_\psi}$ has some analogs of Luzin's $N$-property.

\begin{lem}
\label{lkr7} {\sl Let $A,B\in T_\psi$ with $\diam(A)>0$, $\diam(B)>0$.
If $\mathcal Z\subset [A,B]$ has $T$-measure zero, then
$\Ha^1(\{\Phi(C):C\in \mathcal Z\})=0$.}
\end{lem}

\pr Recall that the Coarea
formula
\begin{equation}\label{Coarea}\int\limits_{E}|\nabla f|\,dx=\int\limits_{\R}\Ha^1(E\cap f^{-1}(y))\,dy
\end{equation}
holds for a measurable set $E$ and the best representative (see
Lemma~\ref{kmpLem1}) of any Sobolev function $f\in
W^{1,1}(\Omega)$ (see, e.g., \cite{Maly}).

Now, let $\mathcal Z\subset [A,B]$ have $T$-measure zero. Set
$E=\cup_{C\in \mathcal Z}C$. Then by definition $\Ha^1(\psi(E))=0$.  Take a Borel set $G\supset \psi(E)$ with $\Ha^1(G)=0$ and put
$\mathcal Z'=\{C\in[A,B]:\psi(C)\in G\}$,  $E'=\cup_{C\in \mathcal Z'}C$.
Then $E'$ is a Borel set as well and $E'\supset E$.
Hence, by Coarea formula~(\ref{Coarea}) applied to $\psi|_{E'}$ we
see that $\nabla\psi(x)=0$ for $\Ha^2$-almost all
$x\in E'$. Then by (\ref{eu0}), $\nabla\Phi(x)=0$ for
$\Ha^2$-almost all $x\in E$. Applying the Coarea formula to $\Phi|_{E'}$, we obtain
$$0=\int_{E'}|\nabla\Phi|\,dx=\int_{\R}\sum_{C\in\mathcal Z'\,:\,\Phi(C)=y}\Ha^1(C)\,\,dy.$$
Since $\Ha^1(C)\ge\min\bigl(\diam(A),\diam(B)\bigr)>0$ for every $C\in
[A,B]$, we have $\Ha^1(\{\Phi(C):C\in \mathcal Z'\})=0$ and this implies the assertion of Lemma~\ref{lkr7}. $\qed$

\medskip

From Lemmas~\ref{kmpLem2.11} and \ref{lkr7} we have

\begin{lemA}
\label{regPhi} If $A,B\in T_\psi$ with $\diam(A)>0$, $\diam(B)>0$,
then $$\Ha^1\bigl(\{\Phi(C):C\in[A,B]\mbox{\rm\ and }C\mbox{\rm\ is not a regular cycle}\}\bigr)=0.$$
\end{lemA}

Denote by $B_0,\dots,B_N$ the elements of $T_\psi$ such that
$B_j\supset \Gamma_j$, $j=0,\dots,N$. By virtue of
Lemma~\ref{lkrSTR}, every element
$C\in[B_i,B_j]\setminus\{B_i,B_j\}$ is a connected component of a
level set of $\psi$ such that the sets $B_i$, $B_j$ lie in
different connected components of~$\R^2\setminus C$.

Put $$\alpha=\max_{j=0,\dots,M}\min\limits_{C\in[B_j,B_N]}\Phi(C).$$
By (\ref{as01}), $\alpha<0$. Take a sequence of positive values $t_i\in(0,-\alpha)$, $i\in\N$, with
$t_{i+1}=\frac12t_i$ and such that the implication
$$
\Phi(C)=-t_i\Rightarrow C\mbox{ is a regular cycle }
$$
holds for every  $ j=0,\dots,M$ and for all $C\in[B_j,B_N]$. The existence of the above sequence follows from Corollary~\ref{regPhi}.

Consider the natural order on the arc
$[C_j,B_N]$, namely, $C'\le C''$ if $C''$ is closer to $B_N$ than
$C'$.  For $j=0,\dots M$ and $i\in\N$ put
$$A^j_i=\max\{C\in[B_j,B_N]:\Phi(C)=-t_i\}.$$
In other words, $A^j_i$ is an~element of the set $\{C\in[B_j,B_N]:\Phi(C)=-t_i\}$
 which is closest to $\Gamma_N$. By construction, each $A^j_i$ is a regular cycle (see Fig.~1 for the case of annulus type domains ($N=1$) ).

Denote by ${V}_{i}$ the connected component of the open set
$\Omega\setminus\bigl(\cup_{j=0}^M A^j_i\bigr)$ such that
$\Gamma_N\subset\partial{V}_{i}$. By construction, the sequence of
domains ${V}_i$ is decreasing, i.e., ${V}_{i}\supset{V}_{i+1}$. Hence, the sequence of sets
$(\partial\Omega)\cap(\partial{V}_{i})$ is nonincreasing:
$$
(\partial\Omega)\cap(\partial{V}_{i})\supseteqq(\partial\Omega)\cap(\partial{V}_{i+1}).
$$
Every set $(\partial\Omega)\cap(\partial{V}_{i})$ consists of several components $\Gamma_l$ with $l>M$ (since arcs $\cup_{j=0}^M A^j_i$ separate
$\Gamma_N$ from $\Gamma_0,\dots,\Gamma_M$, but not necessary
from other $\Gamma_l$\,). Since there are only finitely many components $\Gamma_l$, we conclude  that  for sufficiently large~$i$ the set
$(\partial\Omega)\cap(\partial{V}_{i})$  is independent of $i$. So we may assume, without loss of generality, that
$(\partial\Omega)\cap(\partial{V}_{i})=\Gamma_{K}\cup\dots\cup
\Gamma_N$, where $K\in\{M+1,\dots,N\}$. Therefore, 
\begin{equation}\label{boundary1}\partial{V}_i=A^0_i\cup\dots\cup
A^M_i\cup\Gamma_{K}\cup\dots\cup \Gamma_N.\end{equation}
From Lemma~\ref{kmpLem2.11} we have the uniform convergence
$\Phi_k|_{A^j_i}\rightrightarrows \Phi(A^j_i)=-t_i$ as $k\to\infty$. Thus for every $i\in\N$ there exists $k_i$ such that for all  $k\ge k_i$ 
\begin{equation}\label{boundary0}\Phi_k|_{A^j_i}< -\frac78t_i,\quad\Phi_k|_{A^j_{i+1}}> -\frac58t_i\quad\forall j=0,\dots,M.\end{equation}
Then
\begin{equation}\label{boundary2}
\forall t\in\bigl[\frac58t_i,\frac78t_i\bigr]\ \forall k\ge k_i\quad \Phi_k|_{A^j_i}< -t,\quad\Phi_k|_{A^j_{i+1}}> -t\quad\forall j=0,\dots,M.
\end{equation}

For $k\ge k_i$ and $t\in[\frac58t_i,\frac78t_i]$
denote by $W_{ik}(t)$ the connected component of the open set $\{x\in V_i\setminus\overline V_{i+1}:\Phi_k(x)>-t\}$ such that $\partial W_{ik}(t)\supset A^0_{i+1}$ and put
$S_{ik}(t)=(\partial W_{ik}(t))\cap V_i \setminus\overline V_{i+1}$. Clearly, $\Phi_k\equiv -t$ on $S_{ik}(t)$.
Since the set $S_{ik}(t)$ cannot separate $A^0_{i+1}$ from $A^j_{i+1}$ for $j=1,\dots M$ (indeed,  by (\ref{boundary1}) applied to $V_{i+1}$ we can join $A^0_{i+1}$ and $A^j_{i+1}$ by arcs
in $V_{i+1}\subset\R^2\setminus S_{ik}(t)$\,), we have in addition $\partial W_{ik}(t)\supset A^j_{i+1}$.
Finally, we get
\begin{equation}\label{boundary3}\partial W_{ik}(t)=S_{ik}(t)\cup A^0_{i+1}\cup\dots\cup
A^M_{i+1}\end{equation}
(see Fig.~1). Since by (E--NS) each $\Phi_k$ belongs to   $W^{2,2}_{\loc}(\Omega)$,
by the Morse-Sard theorem for Sobolev functions (see Theorem~\ref{kmpTh1.1}) we have that for almost all $t\in[\frac58t_i,\frac78t_i]$ the level set
$S_{ik}(t)$ consists of finitely many $C^1$-cycles and $\Phi_k$ is differentiable (in classical sense) at every point~$x\in S_{ik}(t)$ with $\nabla\Phi_k(x)\ne0$. The values $t\in[\frac58t_i,\frac78t_i]$ having
the above property will be called $(k,i)$-{\it regular}.  By construction,
\begin{equation}\label{lac-2}
\int_{S_{ik}(t)}\nabla\Phi_k\cdot{\bf
n}\,ds=-\int_{S_{ik}(t)}|\nabla\Phi_k|\,ds<0,
\end{equation}
where $\n$ is the unit outward (with respect to $W_{ik}(t)$) normal vector
to $\partial W_{ik}(t)$.

For $h>0$ denote
$\Gamma_h=\{x\in\Omega:\dist(x,\Gamma_K\cup\dots\cup\Gamma_N)=h)\}$,
$\Omega_h=\{x\in\Omega:\dist(x,\Gamma_K\cup\dots\cup\Gamma_N)<h)\}$.
By elementary results of analysis, there is a constant
$\delta_0>0$ such that for each $h\le\delta_0$ the set $\Gamma_h$
is a union of $N-K+1$ \ $C^1$-smooth  curves homeomorphic to the
circle, and
\begin{equation}\label{lac0.1}
\Ha^1(\Gamma_h)\le C_0\quad\forall h\in(0,\delta_0],
\end{equation}
where $C_0=3\Ha^1(\Gamma_K\cup\dots\cup\Gamma_N)$ is independent of~$h$.

\begin{center}
\includegraphics[scale=0.5]{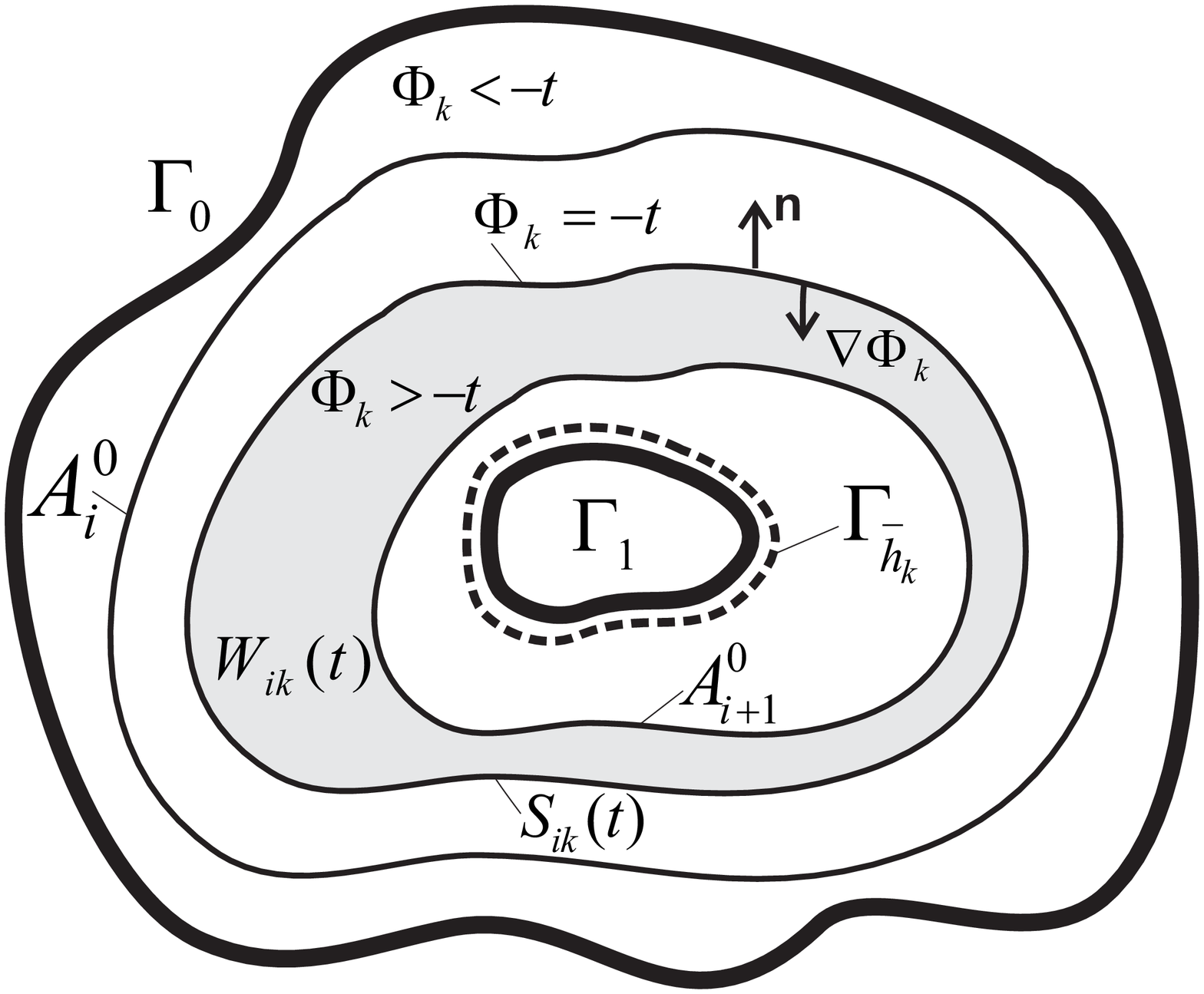}
\end{center}
\begin{center}
Figure 1. The case of an annulus--like domain ($N=1$).
\end{center}

Since $\Phi\neq \const$ on $V_i$, by \eqref{eu0} we have
$\int\limits_{{V}_i}\omega^2\,dx>0$ for each
$i$. Hence, from the weak convergence $\omega_k\rightharpoonup\omega$ in $L^2(\Omega)$  it follows
\begin{lem}
\label{lkr8} {\sl For any $i\in\N$ there exist constants $\e_i>0$,
$\delta_i\in(0,\delta_0)$ and  $k'_i\in\N$ such that
$\int\limits_{{V}_{i+1}\setminus
\Omega_{\delta_i}}\omega_k^2\,dx>\varepsilon_i$ for all $k\ge
k'_i$.}
\end{lem}

The key step is the following estimate.

\begin{lem}
\label{lkr11}{\sl For  any $i\in \N$ there exists $k(i)\in\N$ such
that  the inequality
\begin{equation}\label{mec}
\int\limits_{ S_{ik}(t)}|\nabla\Phi_k|\,ds<\F t
\end{equation}
holds for every $k\ge k(i)$ and for almost all
$t\in[\frac58t_i,\frac78t_i]$, where the constant $\F$ is independent of $t,k$ and $i$. }
\end{lem}

\pr Fix $i\in\mathbb N$ and assume $k\ge k_i$ (see (\ref{boundary0})\,).
Take a sufficiently small $\sigma>0$ (the exact value of $\sigma$
will be specified
 below). We choose the parameter $\delta_\sigma\in(0,\delta_i]$ (see Lemma~\ref{lkr8}) small enough to
satisfy the following conditions:
\begin{equation}\label{lac0.2.1}
\Omega_{\delta_\sigma}\cap A^j_i=\Omega_{\delta_\sigma}\cap
A^j_{i+1}=\emptyset,\quad j=0,\dots,M,
\end{equation}
\begin{equation}\label{lac0.2}
\int\limits_{\Gamma_h}\Phi^2\,ds<\frac13\sigma^2\quad\forall
h\in(0,\delta_\sigma],
\end{equation}
\begin{equation}\label{lac0.3}
-\frac13\sigma^2<\int\limits_{\Gamma_{h'}}\Phi_k^2\,ds-\int\limits_{\Gamma_{h''}}\Phi_k^2\,ds<\frac13\sigma^2
\quad\forall h',h''\in(0,\delta_\sigma]\ \ \forall k\in\mathbb N.
\end{equation}
The last estimate follows from the fact that  for any $q\in(1,2)$ the norms
$\|\Phi_k\|_{W^{1,q}(\Omega)}$ are uniformly bounded. Consequently, the norms $\|\Phi_k\nabla\Phi_k\|_{L^{q}(\Omega)}$ are uniformly bounded as well. In particular, for $q=6/5$ we have
$$\begin{array}{lcr}\displaystyle\biggl|\int\limits_{\Gamma_{h'}}\Phi_k^2\,ds-\int\limits_{\Gamma_{h''}}\Phi_k^2\,ds\biggr|\le
2\int_{\Omega_{h''}\setminus\Omega_{h'}}|\Phi_k|\cdot|\nabla\Phi_k|\,dx
\\[20pt]
\displaystyle
\le 2\biggl(\int_{\Omega_{h''}\setminus\Omega_{h'}}|\Phi_k\nabla\Phi_k|^{6/5}\,dx\biggr)^{\frac56}\meas(\Omega_{h''}\setminus\Omega_{h'})^{\frac16}
\to0\quad\mbox{ as }h',h''\to0.\end{array}$$

From the weak convergence $\Phi_k\rightharpoonup\Phi$ in the space
$W^{1,q}(\Omega)$, $q\in (1,2)$,  it follows    that $\Phi_{k}|_{\Gamma_h}\rightrightarrows\Phi|_{\Gamma_h}$  as $k\to\infty$ for almost all $h\in
(0,\delta_\sigma)$(see
\cite{Amick}, \cite{kpr}\footnote{In \cite{Amick} Amick proved the uniform convergence $\Phi_k\rightrightarrows\Phi$ on almost all
circles. However, his method can be easily modified to    prove the uniform convergence on almost all level lines of every $C^1$-smooth function with nonzero gradient. Such modification was done in
the proof of Lemma~3.3 of \cite{kpr}.})
From the last fact and (\ref{lac0.2})--(\ref{lac0.3}) we see
that there exists $k'\in\mathbb N$ such that
\begin{equation}\label{lac0.4}
\int\limits_{\Gamma_{h}}\Phi_k^2\,ds<\sigma^2
\quad\forall h\in(0,\delta_\sigma]\ \ \forall k\ge k'.
\end{equation}

Obviously, for a function $g\in W^{2,2}(\Omega)$ and for an~arbitrary $C^1$-cycle $S\subset\Omega$ we  have 
$$\int_{S}\nabla^\bot g\cdot{\bf n}\,ds=\int_{S}\nabla g\cdot{\bf l}\,ds=0,$$
where ${\bf l}$ is the tangent vector to~$S$. Consequently, by (\ref{eu0}),
$$\int_{S}\nabla\Phi_k\cdot{\bf n}\,ds=\int_{S}\omega_k\ue^\bot_k\cdot{\bf n}\,ds$$
(recall, that by our assumptions $\fe=\nabla^\bot b$\,).

Now, fix a sufficiently small $\varepsilon>0$ (the exact value of
$\e$ will be specified below). For a given sufficiently large
$k\ge k'$ we make a special procedure to find a number $\bar
h_k\in(0,\delta_\sigma)$ such that the estimates
\begin{equation}\label{lac0.5}
\biggr|\int_{\Gamma_{\bar h_k}}\nabla\Phi_k\cdot{\bf
n}\,ds\biggr|=\biggr|\int_{\Gamma_{\bar
h_k}}\omega_k\ue^\bot_k\cdot{\bf n}\,ds\biggr|<\e,
\end{equation}
\begin{equation}\label{lac0.6}
\int_{\Gamma_{\bar h_k}}|\ue_k|^2\,ds<C_2(\e)\nu_k^2
\end{equation}
hold, where the constant  $C_2(\e)$ {\bf is independent of $k$ and $\sigma$}. To this end
define a sequence of numbers $0=h_0<h_1<h_2<\dots$ by
the recurrent formulas
\begin{equation}\label{lac1}
\int\limits_{U_{j}}|\nabla \ue_k|\cdot|\ue_k|\,dx=\nu_k^2,
\end{equation}
where
$U_j=\{x\in\Omega:\dist(x,\Gamma_K\cup\dots\cup\Gamma_N)\in(h_{j-1},h_{j})\}$.

Since $\int_{\partial
\Omega}|\ue_k|^2\,ds=\frac{(\lambda_k\nu_k)^2}{\nu^2}\|{\bf
a}\|^2_{L^2(\partial\Omega)}$, where $\lambda_k\in(0,1]$, from
(\ref{lac1})  we deduce by induction that
\begin{equation}\label{lac2}
\int_{\Gamma_h}|\ue_k|^2\,ds\le
Cj\nu_k^2 \quad \forall h\in(h_{j-1},h_{j}),
\end{equation}
where $C$ is independent of  $k,j,\sigma$. Consequently,
\begin{equation}\label{lac3}
\int\limits_{U_{j}}|\ue_k|^2\,dx\le(h_{j}-h_{j-1})Cj\nu_k^2.
\end{equation}
Using this estimates and applying the H\"older inequality
to~(\ref{lac1}), we obtain
\begin{equation}\label{lac4}
\nu_k^2=\int\limits_{U_{j}}|\nabla \ue_k|\cdot|\ue_k|\,dx\le
\sqrt{(h_j-h_{j-1})Cj\nu_k^2}\biggl(\int\limits_{U_{j}}|\nabla
\ue_k|^2\,dx\biggr)^{\frac12}.
\end{equation}
Squaring both sides of the last inequality, we have
\begin{equation}\label{lac5}
\frac{\nu_k^2}{h_{j}-h_{j-1}}\le C\,j\int\limits_{U_{j}}|\nabla
\ue_k|^2\,dx.
\end{equation}
We  define  $h_j$ for
$j=1,\dots,j_{max}$, where $j_{\max}$ is the first index satisfying
at least one of the following two conditions.

\medskip
{\sc Stop case 1.} $h_{j_{\max}-1}<\delta_\sigma$,
$h_{j_{\max}}\ge\delta_\sigma$,

\noindent or

\medskip
{\sc Stop case 2.} $Cj_{\max}\int\limits_{U_{j_{\max}}}|\nabla
\ue_k|^2\,dx<\varepsilon$.

\medskip
By construction, $\int_{U_{j}}|\nabla
\ue_k|^2\,dx\ge\frac1{Cj}\varepsilon$ for every $j<j_{\max}$ (since for $j<j_{max}$ the conditions of both Stop cases fail).
Hence,
$$
2\ge\int\limits_{U_{1}\cup\dots\cup U_{j_{\max}-1}}|\nabla
\ue_k|^2\,dx\ge\frac\e{C}\bigl(1+\frac12+\dots+\frac1{j_{\max}-1}\bigr)>C'\e\ln(j_{\max}-1).$$
Consequently, for {\bf both} stop cases we  have the following uniform
estimate
\begin{equation}\label{lac6}
j_{\max}\le 1+ \exp(\frac1{C'\e})
\end{equation}
with $C'$ independent of $k$ and $\sigma$.

Let us  describe the choice of the required distance $\bar h_k$ for
both cases.

Assume that Stop case~1 arises.  Then
$$
\Omega_{\delta_\sigma}\subset U_1\cup\dots\cup U_{j_{max}}
$$
and  by construction (see
(\ref{lac1})--(\ref{lac2})\,) we have
\begin{equation}\label{lac7}
\int\limits_{\Omega_{\delta_\sigma}}|\nabla
\ue_k|\cdot|\ue_k|\,dx\le j_{\max}\nu_k^2,
\end{equation}
\begin{equation}\label{lac8}
\int_{\Gamma_h}|\ue_k|^2\,ds\le
Cj_{\max}\nu_k^2\quad \forall h\in(0,\delta_\sigma].
\end{equation}
From (\ref{lac7}) it follows that there exists $\bar
h_k\in(0,\delta_\sigma)$ such that
\begin{equation}\label{lac10}
\int_{\Gamma_{\bar h_k}}|\nabla
\ue_k|\cdot|\ue_k|\,ds<\frac1{\delta_\sigma}j_{\max}\nu_k^2.
\end{equation}
Then, taking into account that $j_{\max}$ does not depend on
$\sigma$ and $k$ (see (\ref{lac6})\,), and that $\nu_k\to0$ as $k\to\infty$,
we obtain the required estimates (\ref{lac0.5})--(\ref{lac0.6}) for
sufficiently large~$k$.

\medskip

Now, let Stop case~2 arises. By definition of this case and by~(\ref{lac5}), we   obtain
\begin{equation}\label{lac11}
\frac1{h_{j_{\max}}-h_{j_{\max}-1}} \int\limits_{U_{j_{\max}}}
|\nabla
\ue_k|\cdot|\ue_k|\,dx=\frac{\nu_k^2}{h_{j_{\max}}-h_{j_{\max}-1}}<\e.
\end{equation}
Therefore, there exists ${\bar
h_k}\in(h_{j_{\max}-1},h_{j_{\max}})$ such that 
(\ref{lac0.5}) holds. Estimate
(\ref{lac0.6}) follows again from (\ref{lac2}) and the fact that
$j_{\max}$ depends on $\e$ only.
So, for any sufficiently large $k$ we have proved the existence of
$\bar h_k\in(0,\delta_\sigma)$ such that
(\ref{lac0.5})--(\ref{lac0.6}) hold.

Now, for $(k,i)$-regular value $t\in[\frac58t_i,\frac78t_i]$
 consider the domain $$\Omega_{i\bar h_k}(t)=W_{ik}(t)\cup\overline V_{i+1}\setminus\overline \Omega_{\bar h_k}.$$ By construction,
$\partial\Omega_{i\bar h_k}(t)=\Gamma_{\bar h_k}\cup S_{ik}(t)$ (see Fig.~1). Integrating the equation
\begin{equation}\label{cle_lap}\Delta\Phi_k=\omega_k^2+\frac1{\nu_k}\div(\Phi_k\ue_k)-\frac{1}{\nu_k}\fe_k\cdot\ue_k\end{equation}
over the domain $\Omega_{i\bar h_k}(t)$, we have
$$
\int_{ S_{ik}(t)}\nabla\Phi_k\cdot{\bf
n}\,ds+\int_{\Gamma_{\bar h_k}}\nabla\Phi_k\cdot{\bf
n}\,ds=\int_{\Omega_{i\bar h_k}(t)}\omega_k^2\,dx-\frac1{\nu_k}\int_{\Omega_{i\bar h_k}(t)}\fe_k\cdot\ue_k\,dx
$$
$$
+\frac{1}{\nu_k}\int_{ S_{ik}(t)}\Phi_k\ue_k\cdot{\bf n}\,ds+\frac1{\nu_k}\int_{\Gamma_{\bar
h_k}}\Phi_k\ue_k\cdot{\bf n}\,ds
$$
\begin{equation}\label{cle_lap1}
=\int_{\Omega_{i\bar h_k}(t)}\omega_k^2\,dx-\frac1{\nu_k}\int_{\Omega_{i\bar h_k}(t)}\fe_k\cdot\ue_k\,dx-t{\lambda_k}\bar\F+\frac1{\nu_k}\int_{\Gamma_{\bar
h_k}}\Phi_k\ue_k\cdot{\bf n}\,ds,\end{equation} where
$\bar\F=\frac1\nu(\F_1+\dots+\F_M)$. In view of (\ref{lac-2}),
(\ref{lac0.5}), we can estimate
$$
\int_{ S_{ik}(t)}|\nabla\Phi_k|\,ds
\le
t\F+\e+\frac1{\nu_k}\int_{\Omega_{i\bar h_k}(t)}\fe_k\cdot\ue_k\,dx-\int_{\Omega_{i\bar h_k}(t)}\omega_k^2\,dx
$$
\begin{equation}\label{cle_lap2}
+\frac1{\nu_k}
\biggl(\int_{\Gamma_{\bar h_k}}\Phi_k^2\,ds\biggr)^{\frac12}
\biggl(\int_{\Gamma_{\bar h_k}}|\ue_k|^2\,ds\biggr)^{\frac12}
\end{equation} with $\F=|\bar\F|$. By definition,
$\frac{1}{\nu_k}\|\fe_k\|_{L^2(\Omega)}=
\frac{\lambda_k\nu_k}{\nu^2}\|\fe\|_{L^2(\Omega)}\to 0$ as $k\to\infty$. Therefore,
$$
\Big|\frac1{\nu_k}\int_{\Omega_{i\bar h_k}(t)}\fe_k\cdot\ue_k\,dx\Big|\le\varepsilon
$$
for sufficiently large $k$. Using inequalities (\ref{lac0.4}),
(\ref{lac0.6}), we obtain 
\begin{equation}\label{cle_lap3}
\begin{array}{lcr}
\displaystyle
\int_{ S_{ik}(t)}|\nabla\Phi_k|\,ds\le
t\F+2\e+\sigma\sqrt{C_2(\e)}-\int_{\Omega_{i\bar h_k}(t)}\omega_k^2\,dx
\\
\displaystyle\le
t\F+2\e+\sigma\sqrt{C_2(\e)}-\int_{{V}_{i+1}\setminus\Omega_{\delta_i}}\omega_k^2\,dx,
\end{array}
\end{equation}
where $C_2(\e)$ is independent of  $k$ and $\sigma$. Choosing $\e=\frac16\e_i$, $\sigma=\frac1{3\,\sqrt{C_2(\e)}}\e_i$,
\, and a sufficiently large $k$, from
Lemma~\ref{lkr8} we obtain
$2\e+\sigma\sqrt{C_2(\e)}-\int_{{V}_{i+1}\setminus\Omega_{\delta_i}}\omega_k^2\,dx\le0$.
Estimate~(\ref{mec}) is proved. $\qed$

\medskip

Now, we receive the required contradiction using the~Coarea formula.

\begin{lem}
\label{lem_Leray_fc} {\sl Assume that  $\Omega\subset\R^2$ is a bounded domain of type \eqref{domain} with
$C^2$-smooth boundary $\partial\Omega$, $\fe\in W^{1,2}(\Omega)$,  and ${\bf a}\in W^{3/2,2}(\partial\Omega)$ satisfies condition \eqref{flux}. 
Then  assumptions (E-NS) and \eqref{as-prev0-0} lead to a contradiction.}
\end{lem}

\pr
For $i\in\N$ and $k\ge k(i)$ (see
Lemma~\ref{lkr11}) put
$$
E_i=\bigcup\limits_{t\in [\frac58t_i,\frac78t_i]} S_{ik}(t).
$$
 By the Coarea formula  \eqref{Coarea} (see also \cite{Maly}), for any integrable
 function $g:E_i\to\R$ the equality
\begin{equation}\label{Coarea_Phi}\int\limits_{E_i}g|\nabla\Phi_k|\,dx=
\int\limits_{\frac78t_i}^{\frac58t_i}\int_{ S_{ik}(t)}g(x)\,d\Ha^1(x)\,dt
\end{equation}
holds. In particular, taking $g=|\nabla\Phi_k|$ and
using~(\ref{mec}), we obtain
\begin{equation}\label{Coarea_Phi2}\int\limits_{E_i}|\nabla\Phi_k|^2\,dx=
\int\limits_{\frac78t_i}^{\frac58t_i}\int_{ S_{ik}(t)}|\nabla\Phi_k|(x)\,d\Ha^1(x)\,dt
\le
\int\limits_{\frac78t_i}^{\frac58t_i}\F t\,dt=
\F't_i^2
\end{equation}
where $\F'=\frac3{16}\F$ is independent of $i$. Now, taking  $g=1$ in
(\ref{Coarea_Phi})  and using the H\"older inequality we
have
\begin{equation}\label{Coarea_Phi3}
\begin{array}{lcr}
\displaystyle
\int\limits_{\frac78t_i}^{\frac58t_i}\Ha^1\bigl( S_{ik}(t)\bigr)\,dt=
\int\limits_{E_i}|\nabla\Phi_k|\,dx
\\
\displaystyle
\le
\biggl(\int\limits_{E_i}|\nabla\Phi_k|^2\,dx\biggr)^{\frac12}
\bigl(\meas (E_i)\bigr)^{\frac12}\le\sqrt{\F'}t_i\bigl(\meas
(E_i)\bigr)^{\frac12}.
\end{array}
\end{equation}
 By construction, for almost all $t\in [\frac58t_i,\frac78t_i]$ the set $S_{ik}(t)$ is a finite union of smooth cycles and
$S_{ik}(t)$ separates $A^j_i$ from $A^j_{i+1}$ for $j=0,\dots,M$. Thus, each set $S_{ik}(t)$
separates $\Gamma_j$ from $\Gamma_N$. In particular,
$\Ha^1(S_{ik}(t))\ge\min\bigl(\diam(\Gamma_j),\diam(\Gamma_N)\bigr)$.
Hence, the left integral in (\ref{Coarea_Phi3}) is greater than
$Ct_i$, where $C>0$ does not depend on $i$. On the other hand,
evidently,
$\meas(E_i)\le\meas\bigl({V}_i\setminus{V}_{i+1}\bigr)\to0$
as $i\to\infty$. The obtained contradiction finishes the proof of Lemma~\ref{lem_Leray_fc}.
$\qed$\\

\subsubsection{The maximum of $\Phi$ is not attained  at $\partial\Omega$}
\label{max_inside_the_plane_domain}

In this subsection we consider the case (b), when \eqref{as-prev-id} holds.
Adding a constant to the pressure, we  assume, without loss of generality, that
\begin{equation}\label{as-prev0-idd} \max\limits_{j=0,\dots,N}\widehat
p_j<\esssup\limits_{x\in\Omega}\Phi(x)=0.
\end{equation}
Denote $\sigma=\max\limits_{j=0,\dots,N}\widehat
p_j<0$.

As in the previous subsection, we  consider the behavior of $\Phi$ on the Kron-rod tree~$T_\psi$. In particular,  Lemmas~\ref{lkr1}--\ref{lkr7} hold.

\begin{lem}\label{mppr1}{\sl There exists $F\in T_\psi$ such that $\diam F>0$, $F\cap\partial\Omega=\emptyset$, and $\Phi(F)>\sigma$.}
\end{lem}

\pr By assumptions,
$\Phi(x)\le\sigma$ for every $x\in\partial\Omega\setminus A_\ve$ and there is a set of a positive measure $E\subset \Omega$ such that $\Phi(x)>\sigma$ at each~$x\in E$.
In virtue of Theorem~\ref{kmpTh2.1} (iii), there exists a straight-line segment $I=[x_0,y_0]\subset\overline\Omega$ with
$I\cap A_\ve=\emptyset$, $x_0\in\partial\Omega$, $y_0\in E$, such that $\Phi|_I$ is a continuous function.
By construction, $\Phi(x_0)\le\sigma$, $\Phi(y_0)\ge\sigma+\delta_0$ with some $\delta_0>0$.
Take a subinterval $I_1=[x_1,y_0]\subset\Omega$ such that $\Phi(x_1)=\sigma+\frac12\delta_0$ and $\Phi(x)\ge\sigma+\frac12\delta_0$ for each
$x\in[x_1,y_0]$. Then by Bernoulli's Law (see Theorem~\ref{kmpTh2.2}) $\psi\ne \const$ on $I_1$. Hence, we can take $x\in I_1$ such that the preimage $\psi^{-1}(\psi(x))$ consists of a finite union of regular cycles
(see Lemma~\ref{kmpLem2.11}). Denote by $F$ the regular cycle containing~$x$. Then by construction
$\Phi(F)\ge\sigma+\frac12\delta_0$ and by definition of regular cycles~$\diam F>0$ and~$F\cap\partial\Omega=\emptyset$.
$\qed$
\\

Fix $F$ from above Lemma and consider the behavior of $\Phi$ on the Kronrod  arcs $[B_j,F]$, $j=0,\dots N$ (recall, that by $B_j$ we denote the elements of
$T_\psi$ such that $\Gamma_j\subset B_j$). The rest part of this subsection is  similar to that of Subsection~\ref{subsub1} with the following difference:  $F$ plays now the role which was played before by $B_N$, and the calculations become  easier since $F$ lies strictly inside~$\Omega$.

By construction, $\Phi(F)>\Phi(B_j)$ for each $j=0,\dots,N$.
So, using Lemmas~\ref{lkr1}--\ref{lkr7} and Corollary~\ref{regPhi} 
we can find a~sequence
of positive numbers $t_i\in(-\Phi(F),-\sigma)$, $i\in\N$, with
$t_{i+1}=\frac12t_i$, and the corresponding regular cycles~$A^j_i\in[B_j,F]$, $j=0,\dots,N$, with
$\Phi(A^j_i)=-t_i$. Denote by $V_i$ the connected component of the set $\Omega\setminus (A^0_i\cup\dots\cup
A^N_i)$ containing~$F$. By construction, $\overline V_i\subset \Omega$, $V_i\subset\overline V_{i+1}$  and
\begin{equation}\label{mpp3}\partial{V}_i=A^0_i\cup\dots\cup
A^N_i.\end{equation}

By definition of regular cycles (see Lemma~\ref{kmpLem2.11}),  we  again obtain estimates~(\ref{boundary0})--(\ref{boundary2}) for $k\ge k_i$. Accordingly,
for $k\ge k_i$ and $t\in[\frac58t_i,\frac78t_i]$ we can define the domain
$W_{ik}(t)$ as a~connected component of the open set $\{x\in V_i\setminus\overline V_{i+1}:\Phi_k(x)>-t\}$
with \begin{equation}\label{boundary3-ax}\partial W_{ik}(t)=S_{ik}(t)\cup A^0_{i+1}\cup\dots\cup
A^N_{i+1},\end{equation}
where the set
$S_{ik}(t)=(\partial W_{ik}(t))\cap V_i \setminus\overline V_{i+1}
\subset\{x\in V_i:\Phi_k(x)=-t\}$ separates $A^0_{i}\cup\dots\cup
A^N_{i}$ from $A^0_{i+1}\cup\dots\cup
A^N_{i+1}$.
By the Morse-Sard theorem (see Theorem~\ref{kmpTh1.1}) applied to $\Phi_k\in W^{2,2}_\loc(\Omega)$, for almost all $t\in[\frac58t_i,\frac78t_i]$ the level set
$S_{ik}(t)$ consists of finitely many $C^1$-cycles. Moreover, by construction,
\begin{equation}\label{lac-2-ax-mpp}
\int_{S_{ik}(t)}\nabla\Phi_k\cdot{\bf
n}\,ds=-\int_{S_{ik}(t)}|\nabla\Phi_k|\,ds<0,
\end{equation}
where $\n$ is the unit outward normal vector
to $\partial W_{ik}(t)$. As before, we call such values $t\in[\frac58t_i,\frac78t_i]$ \ {\it$(k,i)$-regular}.

Since $\Phi\neq \const$ on $V_i$, from  \eqref{eu0} it follows that
$\int\limits_{{V}_i}\omega^2\,dx>0$ for each
$i$, and taking into account  the weak convergence $\omega_k\rightharpoonup\omega$ in $L^2(\Omega)$  we get
\begin{lem}
\label{lkr8-mpp} {\sl For every $i\in\N$ there exist constants $\e_i>0$,
$\delta_i\in(0,\delta_0)$ and $k'_i\in\N$ such that
$\int\limits_{{V}_{i+1}}\omega_k^2\,dx>\varepsilon_i$ for all $k\ge
k'_i$.}
\end{lem}

Now, we can prove

\begin{lem}
\label{lem_Leray_fc-idd} {\sl Assume that  $\Omega\subset\R^2$ is a bounded domain of type \eqref{domain} with
$C^2$-smooth boundary $\partial\Omega$, ${\bf f}\in W^{1,2}(\Omega)$,  and ${\bf a}\in W^{3/2,2}(\partial\Omega)$ satisfies condition \eqref{flux}.
Then   assumptions (E-NS) and \eqref{as-prev-id} lead to a contradiction.}
\end{lem}

\pr
The proof of this Lemma is similar to that of  Lemma~\ref{lkr11}. However,  the situation now is  more easy, since we separate $V_i$ from the whole boundary~$\partial\Omega$.
Fix $i\in\mathbb N$ and assume that $k\ge k_i$ (see (\ref{boundary0})\,).
For a $(k,i)$-regular value $t\in[\frac58t_i,\frac78t_i]$
 consider the domain $$\Omega_{ik}(t)=W_{ik}(t)\cup\overline V_{i+1}.$$ By construction,
$\partial\Omega_{ik}(t)=S_{ik}(t)$. Integrating identity~(\ref{cle_lap})
over  $\Omega_{ik}(t)$, we obtain
\begin{equation}\label{mppprr}
\begin{array}{lcr}\displaystyle
0>\int_{ S_{ik}(t)}\nabla\Phi_k\cdot{\bf
n}\,ds=\int_{\Omega_{ik}(t)}\omega_k^2\,dx+\frac{1}{\nu_k}\int_{ S_{ik}(t)}\Phi_k\ue_k\cdot{\bf n}\,ds\\[15pt]\displaystyle-\frac{1}{\nu_k}\int_{\Omega_{ik}(t)}\fe_k\cdot \ue_k\,dx
=\int_{\Omega_{ik}(t)}\omega_k^2\,dx-\frac{t}{\nu_k}\int_{S_{ik}(t)}\ue_k\cdot{\bf n}\,ds\\[15pt]\displaystyle-\frac{1}{\nu_k}\int_{\Omega_{ik}(t)}\fe_k\cdot \ue_k\,dx
=\int_{\Omega_{ik}(t)}\omega_k^2\,dx-\frac{1}{\nu_k}\int_{\Omega_{ik}(t)}\fe_k\cdot \ue_k\,dx,\end{array}\end{equation}
and, as before, we have a contradiction with Lemma~\ref{lkr8-mpp}.
 $\qed$
\\

{\bf Proof of Theorem \ref{kmpTh4.1}.} Let the hypotheses of Theorem \ref{kmpTh4.1} be satisfied. Suppose that its assertion fails. Then, by Lemma~\ref{lem_Leray}, there exist $ \ve, p$ and a sequence $(\ue_k,p_k)$ satisfying (E-NS), and by Lemmas~\ref{lem_Leray_fc-idd} and \ref{lem_Leray_fc} these assumptions lead to a contradiction. $\qed$

\section{Axially symmetric case}
\label{Axial}
\setcounter{theo}{0} \setcounter{lem}{0}
\setcounter{lemr}{0}\setcounter{equation}{0}

First, let us specify some notations. Let $O_{x_1},O_{x_2},O_{x_3}$
be coordinate axis in $\R^3$  and $\theta=\arctg(x_2/x_1)$,
$r=(x_1^2+x_2^2)^{1/2}$, $z=x_3$ be cylindrical coordinates.
Denote by $v_\theta,v_r,v_z$ the projections of the vector ${\bf
v}$ on the axes $\theta,r,z$.

A function $f$ is said to be {\it axially symmetric} if it does
not depend on~$\theta$. A vector-valued function ${\bf
h}=(h_r,h_\theta,h_z)$ is called {\it axially symmetric} if
$h_r$, $h_\theta$  and $h_z$ do not depend on~$\theta$. A
vector-valued function ${\bf h}'=(h_r,h_\theta,h_z)$ is called
{\it axially symmetric without rotation} if $h_\theta=0$ while
$h_r$ and $h_z$ do not depend on~$\theta$.

The main result of this section is as follows.

\begin{theo} \label{kmpTh4.ax} {\sl Assume that  $\Omega\subset\R^3$
is a bounded axially symmetric domain of type (\ref{domain})
with $C^2$-smooth boundary $\partial\Omega$. If $\fe\in W^{1,2}(\Omega)$, ${\bf a}\in W^{3/2,2}(\partial\Omega)$
are axially symmetric and ${\bf a}$ satisfies  condition $(\ref{flux})$, then $(\ref{NS})$ admits at least one weak axially
symmetric solution. Moreover, if ${\fe}$ and $\bf a$ are axially symmetric
without rotation, then $(\ref{NS})$ admits at least one weak
  axially
symmetric solution without rotation.}
\end{theo}

Using the   ``reductio ad absurdum'' Leray argument (the main idea is
presented in  Section~\ref{Leray_arg} for the plane case;
specific details concerning the axially symmetric case can be found in
\cite{kpr_a_arx}), it is possible to prove the following

 \begin{lem}
\label{lem_Leray_symm} {\sl Assume that  $\Omega\subset\R^3$ is a bounded axially symmetric domain of type \eqref{domain} with
$C^2$-smooth boundary $\partial\Omega$,  $\fe=\curl\,{\bf b}$, ${\bf b}\in W^{2,2}(\Omega)$, ${\bf a}\in W^{3/2,2}(\partial\Omega)$
are axially symmetric, and ${\bf a}$ satisfies  condition $(\ref{flux})$.  If the assertion
of Theorem~\ref{kmpTh4.ax} is false, then there exist $ \ve, p$ with the following properties.

\medskip

(E-AX) \ \,The axially
symmetric functions $\ve\in W^{1,2}(\Omega)$, $p\in
W^{1,3/2}(\Omega)$ satisfy the Euler system \eqref{2.1}.

\medskip

(E-NS-AX) \ \,Condition (E-AX) is satisfied and there exist a
sequences of axially symmetric functions $\ue_k \in
{W^{1,2}(\Omega)}$, $p_k\in {W^{1,q}(\Omega)}$ and numbers
$\nu_k\to0+$, $\lambda_k\to\lambda_0>0$ such that the norms
$\|\ue_k\|_{W^{1,2}(\Omega)}$, $\|p_k\|_{W^{1,3/2}(\Omega)}$ are
uniformly bounded,  the pair $(\ue_k,p_k)$ satisfies \eqref{NSk}
with $\fe_k=\frac{\lambda_k\nu_k^2}{\nu^2}\,{\bf f}$, \ ${\bf a}_k=\frac{\lambda_k\nu_k}\nu\,{\bf a}$,  and
\begin{equation}
\label{E-NS-ax}
\|\nabla\ue_k\|_{L^2(\Omega)}\to1,\qquad \ue_k\rightharpoonup \ve\mbox{ \ in \
}W^{1,2}(\Omega),\qquad p_k\rightharpoonup p\mbox{ \ in \
}W^{1,3/2}(\Omega).
\end{equation}
Moreover, $\ue_k\in W^{3,2}_{\loc}(\Omega)$ and $p_k\in W^{2,2}_{\loc}(\Omega)$.
}
\end{lem}

As in the previous section, in order to prove existence Theorem~\ref{kmpTh4.ax},  we need to show that  conditions (E-NS-AX) lead to a contradiction.

\medskip

Assume that
$$\Gamma_j\cap O_{x_3}\ne\emptyset,\quad j=0,\dots,M',$$
$$\Gamma_j\cap O_{x_3}=\emptyset,\quad j=M'+1,\dots,N.$$

Let $P_+=\{(0,{x_2},{x_3}):{x_2}>0,\ {x_3}\in\R\}$, \
${\D}=\Omega\cap P_+$. Obviously, on
$P_+$ the coordinates $x_2,x_3$ coincide with the coordinates $r,z$.

For a set $A\subset \R^3$ put $\breve{A}:=A\cap P_+$, and for  $B\subset P_+$ denote
by $\widetilde B$ the set in $\R^3$ obtained by rotation of $B$ around $O_z$-axis.

One can easily see that
\\

(S${}_1$) ${\D}$ is a bounded plane domain with Lipschitz
boundary. Moreover, $\breve\Gamma_j$ is a connected set for every
$j=0,\dots,N$. In other words,
$\{\breve\Gamma_j:j=0,\dots,N\}$ coincides with the family of all
connected components of the set $P_+\cap\partial{\D}$.
\\

Hence, ${\bf v}$ and $p$ satisfy the following system in the plane
domain~${\mathcal D}$:
\begin{equation} \label{2.1'}
\left\{\begin{array}{rcl}
\dfrac{\partial p}{\partial
r}-\dfrac{(v_\theta)^2}r+v_r\dfrac{\partial v_r}{\partial r}+
v_z\dfrac{\partial v_r}{\partial z}=0,\\[9pt]
\dfrac{\partial p}{\partial
z}+v_r\dfrac{\partial v_z}{\partial r}+ v_z\dfrac{\partial
v_z}{\partial z}=0,\\[9pt]
\dfrac{v_\theta v_r}r+v_r\dfrac{\partial v_\theta}{\partial r}+
v_z\dfrac{\partial v_\theta}{\partial z}=0,\\[9pt]
\dfrac{\partial (rv_r)}{\partial r}+\dfrac{\partial
(rv_z)}{\partial z}=0
\end{array}\right.
\end{equation}
(these equations are satisfied for almost all $x\in{\mathcal
D}$\,) and
\begin{equation} \label{zbc}
\ve(x)=0\quad\mbox{ for }\Ha^1\mbox{-almost all }x\in P_+\cap\partial\D.
\end{equation}

We have the following integral estimates: ${\bf v}\in
W^{1,2}_\loc({\mathcal D})$,
\begin{equation}
\label{ax'3}\int_{{\mathcal D}}r|\nabla{\bf
v}(r,z)|^2\,drdz<\infty,
\end{equation}
and, by the Sobolev embedding theorem for three--dimensional domains, \linebreak$\ve\in L^6(\Omega)$, i.e.,
\begin{equation}
\label{un_ax3}\int_{{\mathcal D}}r|{\bf v}(r,z)|^6\,drdz<\infty.
\end{equation}
 Also, the condition $\nabla p\in
L^{3/2}(\Omega)$ can be written as
\begin{equation}
\label{un_p2}\int_{{\mathcal D}}r|\nabla
p(r,z)|^{3/2}\,drdz<\infty.
\end{equation}

\subsection{Some previous results on Euler equations}
\label{EAXprev}

The next statement  was proved in \cite[Lemma 4]{KaPi1} and in
\cite[Theorem 2.2]{Amick}.

\begin{theo}
\label{kmpTh2.3''} {\sl If conditions {\rm (E-AX)} are
satisfied, then
\begin{equation} \label{bp2-ax} \forall j\in\{0,1,\dots,N\} \ \exists\, \widehat p_j\in\R:\quad
p(x)\equiv \widehat p_j\quad\mbox{for }\Ha^2-\mbox{almost all }
x\in\Gamma_j.\end{equation} In particular, by axial symmetry,
\begin{equation}
\label{bp1}p(x)\equiv \widehat p_j\quad\mbox{for }\Ha^1-\mbox{almost all }
x\in \breve\Gamma_j.\end{equation} }
\end{theo}

The following result was obtained  in~\cite{kpr_a_arx}.

\begin{theo}
\label{kmpTh2.3_un} {\sl If conditions {\rm (E-AX)} are satisfied, then $\widehat p_0=\dots=\widehat p_{M'}$, where
$\widehat p_j$ are the constants from Theorem~\ref{kmpTh2.3''}.}
\end{theo}

We need a weak version of Bernoulli's law for a Sobolev solution
$({\bf v}, p)$ to the Euler equations \eqref{2.1'} (see
Theorem~\ref{kmpTh2.2-ax} below).

From the last equality in (\ref{2.1'}) and from~(\ref{ax'3}) it
follows that there exists a stream function $\psi\in
W^{2,2}_{\loc}({\mathcal D})$ such that
\begin{equation}
\label{ax7}\frac{\partial\psi}{\partial
r}=-rv_z,\quad\frac{\partial\psi}{\partial z}=rv_r.
\end{equation}

Fix a point $x_*\in{\mathcal D}$. For $\varepsilon>0$ denote by
$\D_\varepsilon$ the connected component of ${\mathcal
D}\cap\{(r,z):r>\varepsilon\}$ containing~$x_*$. Since
\begin{equation}
\label{axc1}\psi\in
W^{2,2}(\D_\varepsilon)\quad\forall\varepsilon>0,
\end{equation}
by Sobolev embedding theorem, $\psi\in C(\bar\D_{\varepsilon})$.
Hence $\psi$ is continuous at points of $\bar{\mathcal
D}\setminus O_z=\bar{\mathcal D}\setminus \{(0,z):z\in\R\}$.

\begin{lem}
\label{lkr-ax-STR} {\rm [cf. Lemma~\ref{lkrSTR}] }{\sl If conditions {\rm (E-AX)} are satisfied, then
there exist constants $\xi_0,\dots, \xi_N\in\R$ such that
$\psi(x)\equiv \xi_j$ on each curve $\breve\Gamma_j$,
$j=0,\dots,N$.}
\end{lem}

\pr In virtue of (\ref{zbc}), (\ref{ax7}), we have $\nabla
\psi(x)=0$ for $\mathfrak{H}^1$-almost all $x\in\partial{\mathcal
D}\setminus O_z$. Then the Morse-Sard property (see
Theorem~\ref{kmpTh1.1}) implies that
$$
\mbox{for \ any\ connected\ set\ }C\subset\partial{\mathcal
D}\setminus O_z\ \exists\, \alpha=\alpha(C)\in\R :\quad
\psi(x)\equiv\alpha\ \forall x\in C.
$$
Hence, since $\breve\Gamma_j$ are connected (see (S${}_1$)\,),  the lemma follows. $\qed$

Denote by $\Phi= p+\dfrac{|{\bf v}|^2}{2}$ the total head pressure
corresponding to the solution $({\bf v}, p)$. Obviously,
\begin{equation}
\label{axc2nloc}\int_{{\mathcal D}}r|\nabla
\Phi(r,z)|^{3/2}\,drdz<\infty.
\end{equation}
Hence,
\begin{equation}
\label{axc2}\Phi\in
W^{1,3/2}(\D_\varepsilon)\quad\forall\varepsilon>0.
\end{equation}

Applying Lemmas~\ref{kmpLem1}, \ref{kmpThDor}, and
Remark~\ref{kmpRem1} to the functions $\ve,\psi,\Phi$ we get the
following

\begin{lem}
\label{kmpTh2.1-ax} {\sl If conditions {\rm (E-AX)} hold, then
there exists a set $A_{\ve}\subset \overline\D$ such that:

 {\rm (i)}\quad $ \mathfrak{H}^1(A_{\ve})=0$;

{\rm (ii)} for  all  $x=(r,z)\in\D\setminus A_{\ve}$
\begin{displaymath}
\lim\limits_{\rho\to
0}\dashint\nolimits_{B_\rho(x)}|\ve(z)-\ve(x)|^2dz=\lim\limits_{\rho\to
0}\dashint\nolimits_{B_\rho(x)}|{\Phi}(z)-{\Phi}(x)|^2dz=0,
\end{displaymath}
moreover, the function $\psi$ is differentiable at $x$ and
$\nabla\psi(x)=(-rv_z(x), rv_r(x))$;

{\rm  (iii) } for every $\varepsilon >0$ there exists a set
$U\subset \mathbb{R}^2$ with
$\mathfrak{H}^1_\infty(U)<\varepsilon$, $A_{\ve}\subset U$, and such that the
functions $\ve, \Phi$ are continuous on $\overline\D\setminus
(U\cup O_z)$.}
\end{lem}

The next two results were obtained in \cite{kpr_a_arx}.

\begin{theo}[Bernoulli's Law]
\label{kmpTh2.2-ax} {\sl Let  conditions {\rm (E-AX)} be valid and let $A_\ve$ be a set from Lemma~\ref{kmpTh2.1-ax}. For any compact connected
set $K\subset \bar\D\setminus O_z$ the
following property holds: if
\begin{equation}
\label{2.4-axx} \psi\big|_{K}=\const,
\end{equation}
then
\begin{equation}
\label{2.5'-axx}  \Phi(x_1)=\Phi(x_2) \quad\mbox{for
 all \,}x_1,x_2\in K\setminus A_{\bf v}.
\end{equation}
}
\end{theo}

We also need the following assertion from \cite{kpr_a_arx} concerning the behavior
of the total head pressure near the singularity axis $O_z$.

\begin{lem} \label{kmp_b}{\sl
Assume that conditions {\rm (E-AX)} are satisfied. Let $K_i$ be
a sequence of compact sets with the following properties:
$K_i\subset\bar{\mathcal D}\setminus O_z$, $\psi|_{K_i}=\const$,
and $\lim\limits_{i\to\infty}\inf\limits_{(r,z)\in K_i}r=0$, \ $\varliminf\limits_{i\to\infty}\sup\limits_{(r,z)\in K_i}r>0$.
Then $\Phi(K_i)\to \widehat p_0$ as
$i\to\infty$. }
\end{lem}

Here we denote by $\Phi(K_i)$ the corresponding constant $c_i\in\R$ such that $\Phi(x)=c_i$ for all $x\in K_i\setminus A_\ve$ (see Theorem~\ref{kmpTh2.2-ax}).

\subsection{Obtaining a contradiction}

We consider three possible cases.

(a) The maximum of $\Phi$ is attained on the boundary component intersecting the symmetry axis:
\begin{equation}\label{as-prev1}
\widehat p_0=\max\limits_{j=0,\dots,N}\widehat p_j=\esssup\limits_{x\in\Omega}\Phi(x).
\end{equation}

(b) The maximum of $\Phi$ is attained on a boundary component
which does not intersect the symmetry axis:
\begin{equation}\label{as1-axxx}
\widehat p_0<\widehat p_N=\max\limits_{j=0,\dots,N}\widehat
p_j=\esssup\limits_{x\in\bar\Omega}\Phi(x),
\end{equation}

(c) The  maximum of $\Phi$ is not attained on $\partial\Omega$:
\begin{equation}\label{as-prev-id-ax} \max\limits_{j=0,\dots,N}\widehat
p_j<\esssup\limits_{x\in\Omega}\Phi(x).
\end{equation}

\subsubsection{The case $\esssup\limits_{x\in\Omega}\Phi(x)=\widehat p_0$.}
\label{EPcontr-axx}

Let us consider  case (\ref{as-prev1}).
Adding a constant to the pressure $p$, we can assume, without loss of
generality, that
\begin{equation}\label{as-prev1'}
\widehat p_0=\esssup\limits_{x\in\Omega}\Phi(x)=0.
\end{equation}

Since the identity $\widehat p_0=\widehat p_1=\dots=\widehat p_N$
is impossible (see Corollary~\ref{ppp1}, which is valid also for the
axial-symmetric case), we have that $\widehat p_j<0$
for some $j\in\{M'+1,\dots,N\}$ (recall, that by Theorem~\ref{kmpTh2.3_un},
$\widehat p_0=\dots=\widehat p_{M'}=0$\,).

Now, we receive a~contradiction following the
arguments of \cite{kpr_a_arx}, \cite{kpr_a_crm}. For  reader's
convenience, we recall these arguments.
From  equation $(\ref{2.1}_1)$ we obtain the identity
\begin{equation}\label{repr}
\begin{array}{lcr} 0=x\cdot\nabla p(x)+x\cdot\big(  {\bf
v}(x)\cdot\nabla \big) {\bf v}(x)\\[4pt]=\div\big[x\,  p(x)+ \big(  {\bf
v}(x)\cdot x\big)  {\bf v}(x)\big]-  p(x)\,\div x-|  {\bf v}(x)|^2
\\[4pt]
=\div\big[x\, p(x)+ \big(  {\bf
v}(x)\cdot x\big)  {\bf v}(x)\big]-3\Phi(x)+\frac12|\ve(x)|^2.
\end{array}
\end{equation}
Integrating it over $\Omega$, we derive
\begin{displaymath}
0\ge\int\limits_\Omega \bigl[3\Phi(x)-\frac12|\ve(x)|^2\bigr]\,dx-=\int\limits_{\partial\Omega}  p(x)
\big(x\cdot{\bf n}\big)\,ds=
\sum\limits_{j=0}^N\widehat p_j\int\limits_{\Gamma_j} \big(x\cdot{\bf
n}\big)\,ds
\end{displaymath}
\begin{displaymath}
=-\sum\limits_{j=1}^N\widehat p_j\int\limits_{\Omega_j}{\rm
div}\,x\,dx=  -3\sum\limits_{j=1}^N\widehat p_j|\Omega_j|>0.
\end{displaymath}
The obtained
contradiction finishes the proof for  case~(\ref{as-prev1}).

\subsubsection{The case $\widehat p_0<\widehat p_N=\esssup\limits_{x\in\bar\Omega}\Phi(x)$.}\label{Euler-contr-2}

Suppose that (\ref{as1-axxx}) holds. We may assume, without loss of generality, that the maximum value is zero, i.e.,
\begin{equation}\label{as1-ax}
\widehat p_0<\widehat p_N=\max\limits_{j=0,\dots,N}\widehat
p_j=\esssup\limits_{x\in\bar\Omega}\Phi(x)=0.
\end{equation}
From Theorem~\ref{kmpTh2.3_un} we have
\begin{equation}\label{pro-ravno}
\widehat p_0=\dots=\widehat p_{M'}<0.
\end{equation}
Change (if necessary) the numbering  of the boundary components
$\Gamma_{M'+1}$, \dots, $\Gamma_{N-1}$ so that
\begin{equation}\label{as01-ax}
\widehat p_j<0, \quad j=0,\dots,M,\quad M\ge M', \end{equation}
\begin{equation}\label{as1-ax'}
\widehat p_{M+1}=\dots=\widehat p_N=0.
\end{equation}

The first goal  is to remove a neighborhood of the singularity line $O_z$ from our considerations.
Then, we can reduce the proof to the plane case considered in Subsection~\ref{subsub1}.

Take $r_0>0$ such that the open set $\D_{\varepsilon}=\{(r,z)\in\D:r>\varepsilon\}$ is connected
for every $\e\le r_0$ (i.e., $\D_\e$ is a domain), and
\begin{equation}\label{ass-ax-bound}\begin{array}{lcr}\displaystyle
\breve\Gamma_j\subset \overline D_{r_0}\quad\mbox{and}\quad\ \inf\limits_{(r,z)\in\breve\Gamma_j}r\ge 2r_0,\quad j={M'}+1,\dots,N,\\[12pt]
\displaystyle
\breve\Gamma_j\cap \overline D_{\e}\mbox{ is a connected set}\\ \mbox{ and }
\sup\limits_{(r,z)\in\breve\Gamma_j\cap \overline D_{\e}}r\ge 2r_0,\quad j=0,\dots, M',\ \e\in(0,r_0].\end{array}
\end{equation}

Let a set
$C\subset \D_\e$  separate $\breve\Gamma_i$ and  $\breve \Gamma_j$ in $\D_\e$, i.e.,  $\breve\Gamma_i\cap\D_\e$ and $\breve\Gamma_j\cap\D_\e$ lie in different connected components of $\overline\D_\e\setminus C$. Obviously, for $\e\in(0,r_0]$ there exists a constant $\delta(\varepsilon)>0$ such that the uniform estimate $\sup\limits_{(r,z)\in C} r\ge \delta(\e)$ holds (see Fig.~2). Moreover, the function $\delta(\e)$ is nondecreasing. In particular,
\begin{equation}\label{acs-ax}
\delta(\varepsilon)\ge\delta(r_0),\quad\e\in(0,r_0].\end{equation}

By Remark~\ref{kmpRem1.2} and Lemma~\ref{lkr-ax-STR}, we can apply Kronrod's
results to the stream function~$\psi|_{\bar\D_{\varepsilon}}$, $\varepsilon\in(0,r_0]$. Accordingly, $T_{\psi,\e}$
means the corresponding Kronrod tree for the restriction
$\psi|_{\bar\D_{\e}}$. Define the total head pressure on~$T_{\psi,\e}$
 as we did in
Subsection~\ref{subsub1}. Then the following analog of
Lemma~\ref{lkr1} holds

\begin{lem}
\label{lkr1-ax} {\sl Let $A,B\in T_{\psi,\e}$, where $\e\in(0,r_0]$, $\diam A>0$, and $\diam B>0$. Consider
the corresponding arc $[A,B]\subset T_{\psi,\e}$ joining $A$ to $B$
$($see Lemmas~\ref{kmpLem6}, \ref{kmpLem7} $)$. Then the
restriction $\Phi|_{[A,B]}$ is a continuous function.}
\end{lem}

The lemma is proved using the argument of Lemma~\ref{lkr1} and taking into account  the~above definitions, Theorem~\ref{kmpTh2.2-ax}, and
the continuity properties of $\Phi$ (see
Lemma~\ref{kmpTh2.1-ax}~(iii))\,).\\

Denote by $B^\e_0,\dots,B^\e_N$ the elements of $T_{\psi,\e}$ such that
$B^\e_j\supset \breve{\Gamma}_j\cap\bar\D_{\e}$, $j=0,\dots,{M'}$,
and $B^\e_j\supset \breve{\Gamma}_j$, $j={M'}+1,\dots,N$. By construction, 
$\Phi(B^\e_j)<0$ for $j=0,\dots,M$, and $\Phi(B^\e_j)=0$ for $j=M+1,\dots,N$.
For $r>0$ let $L_r$ be the horizontal straight line $L_r=\{(r,z):z\in\R\}$. We have

\begin{lem}\label{(*)}
{\sl There exist ${r_*}\in(0,r_0]$ and $C_j\in[B^{{r_*}}_j,B^{{r_*}}_N]$, $j=0,\dots, M$, such that
$\Phi(C_j)<0$ and $C\cap L_{{r_*}}=\emptyset$ for all $C\in [C_j,B^{{r_*}}_N]$. }
\end{lem}
\pr
Suppose that the lemma fails for some~$j=0,\dots,M$. Then it is easy to construct $r_i\to0$ and $C^i\in[B^{{r_i}}_j,B^{{r_i}}_N]$ such that $C^i\cap L_{r_i}\ne\emptyset$ and $\Phi(C^i)\to0$.
Since by (\ref{as01-ax}) $\widehat p_0<0$, we have $\Phi(C^i)\nrightarrow \widehat p_0$. By (\ref{acs-ax}), $\sup\limits_{(r,z)\in C^i}r\ge \delta(r_0)$.
Therefore, we have a contradiction with Lemma~\ref{kmp_b}, and the result is proved. $\qed$
\\

Lemma~\ref{(*)} allows us to remove a neighborhood of the
singularity line $O_z$ from our argument. Thus, we can apply
the approach developed in Subsection~\ref{subsub1} for the plane case.
Put, for simplicity,  $T_\psi=T_{\psi,{r_*}}$ and  $B_j=B^{{r_*}}_j$.
Since $\partial \D_{{r_*}}\subset B_0\cup\dots\cup B_N\cup L_{{r_*}}$ and the set $\{B_0,\dots,B_N\}\subset T_\psi$ is finite, we can change $C_j$ (if necessary) so that the assertion of Lemma~\ref{(*)}   takes the following stronger form:
 \begin{equation}\label{lkr2-ax'"}
\displaystyle\forall j=0,\dots,M\ \  C_j\in [B_j,B_N],\ \ \Phi(C_j)<0,
\end{equation}
and
\begin{equation}\label{lkr2-ax'}
\displaystyle C\cap \partial\D_{r_*}=\emptyset\quad\forall C\in [C_j,B_N].
\end{equation}

 Observe that $\Gamma_j\cap L_{r_*}\ne\emptyset$ for $j=0,\dots,M'$. Therefore, if a cycle $C\in T_\psi$ separates $\Gamma_N$ from $\Gamma_0$ and $C \cap\partial\D_{r_*}=\emptyset$, then $C$ separates $\Gamma_N$ from $\Gamma_j$ for all $j=1,\dots,M'$. So we can take $C_0=\dots=C_{M'}$ (see Fig.2) and to consider only the Kronrod arcs $[C_{M'}, B_N]$, $\dots$, $[C_{N}, B_N]$.

\begin{center}
\includegraphics[scale=0.4]{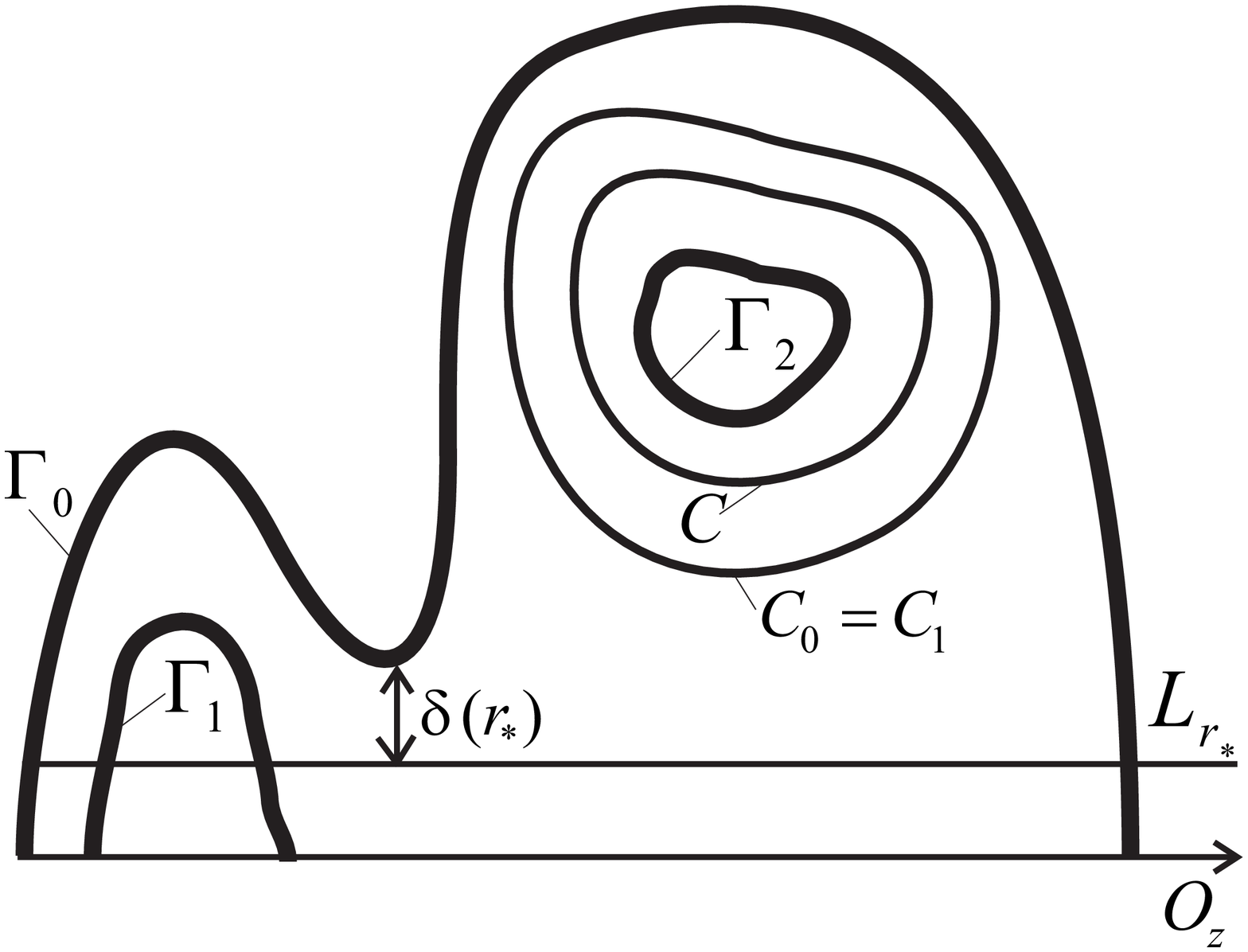}
\end{center}
\begin{center}
Figure 2. The domain $\D$ for the case $M'=1, N=2$.
\end{center}

Recall that a set $\mathcal Z\subset T_\psi$ has $T$-measure
zero, if $\Ha^1(\{\psi(C):C\in \mathcal Z\})=0$.

\begin{lem}
\label{regc-ax} {\sl For every $j=M',\dots,M $,  $T$-almost
all $C\in[C_j,B_N]$ are $C^1$-curves homeomorphic to
the circle. Moreover, all the functions $\Phi_k|_C$
are continuous and the sequence $\{\Phi_k|_C\}$ converges to
$\Phi|_C$ uniformly: $\Phi_k|_C\rightrightarrows\Phi|_C$.}
\end{lem}

 The first assertion of the lemma follows from
Theorem~\ref{kmpTh1.1}~(iv) and ~(\ref{lkr2-ax'}). The validity of the second one
for $T$-almost all
$C\in[C_j,B_N]$ was proved in~\cite[Lemma 3.3]{kpr}. \\

Below we will call {\it regular } the cycles $C$ which satisfy the~assertion of Lemma~\ref{regc-ax}.

From Lemmas~\ref{regc-ax} and \ref{lkr7} (which is also valid for the~axially symmetric case) we obtain

\begin{lemA}
\label{regPhi-ax} For each $j=M',\dots,M$, we have
$$\Ha^1\bigl(\{\Phi(C):C\in[C_j,B_N]\mbox{\rm\ and }C\mbox{\rm\ is not a regular cycle}\}\bigr)=0.$$
\end{lemA}

As  in the plane case (see Subsection~\ref{subsub1}), we can take a~sequence
of positive values $t_i$  with
$t_{i+1}=\frac12t_i$, the corresponding regular cycles~$A^j_i\in[C_j,B_N]$ with
$\Phi(A^j_i)=-t_i$, and the sequence of domains~${V}_{i}\subset\D_{r_*}$ with
\begin{equation}\label{boundary1-ax}\partial{V}_i=A^{M'}_i\cup\dots\cup
A^M_i\cup\breve\Gamma_{K}\cup\dots\cup \breve\Gamma_N,\end{equation}
where $K\ge M+1$ is independent of~$i$.

By the definition of regular cycles, we have again  estimates~(\ref{boundary0})--(\ref{boundary2}) for $k\ge k_i$. Accordingly,
for $k\ge k_i$ and $t\in[\frac58t_i,\frac78t_i]$ we can define the domain
$W_{ik}(t)$ as a~connected component of the open set $\{x\in V_i\setminus\overline V_{i+1}:\Phi_k(x)>-t\}$
with \begin{equation}\label{boundary3-ax-wa}\partial W_{ik}(t)=S_{ik}(t)\cup A^{M'}_{i+1}\cup\dots\cup
A^M_{i+1},\end{equation}
where the set
$S_{ik}(t)=(\partial W_{ik}(t))\cap V_i \setminus\overline V_{i+1}
\subset\{x\in V_i:\Phi_k(x)=-t\}$ separates $A^{M'}_{i}\cup\dots\cup
A^M_{i}$ from $A^{M'}_{i+1}\cup\dots\cup
A^M_{i+1}$. Since $\Phi_k\in W^{2,2}_\loc(\Omega)$ (see (E-NS-AX)\,), by the Morse-Sard theorem (see Theorem~\ref{kmpTh1.1}),
for almost all $t\in[\frac58t_i,\frac78t_i]$ the level set
$S_{ik}(t)$ consists of finitely many $C^1$-cycles and $\Phi_k$ is differentiable (in classical sense) at every point $x\in S_{ik}(t)$ with $\nabla\Phi_k(x)\ne0$.
Therefore, $\widetilde S_{ik}(t)$ is a finite union of smooth surfaces (tori), and by construction,
\begin{equation}\label{lac-2-ax}
\int_{\widetilde S_{ik}(t)}\nabla\Phi_k\cdot{\bf
n}\,dS=-\int_{\widetilde S_{ik}(t)}|\nabla\Phi_k|\,dS<0,
\end{equation}
where $\n$ is the unit outward normal vector
to $\partial\widetilde W_{ik}(t)$ (recall, that for a set $B\subset P_+$ we denote
by $\widetilde B$ the set in $\R^3$ obtaining by rotation of $B$ around $O_z$-axis).

For $h>0$ denote
$\Gamma_h=\{x\in\Omega:\dist(x,\Gamma_K\cup\dots\cup\Gamma_N)=h)\}$,
$\Omega_h=\{x\in\Omega:\dist(x,\Gamma_K\cup\dots\cup\Gamma_N)<h)\}$.
Since the distance function $\dist(x,\partial\Omega)$ is $C^1$--regular and the norm of its gradient is equal to one
in the neighborhood of $\partial\Omega$, there is a constant
$\delta_0>0$ such that for every $h\le\delta_0$ the set $\Gamma_h$
is a union of $N-K+1$ \ $C^1$-smooth  surfaces homeomorphic to the
torus, and
\begin{equation}\label{lac0.1-ax}
\Ha^2(\Gamma_h)\le c_0\quad\forall h\in(0,\delta_0],
\end{equation}
where the constant $c_0=3\Ha^2(\Gamma_K\cup\dots\cup\Gamma_N)$ is independent 
of~$h$.

By a direct calculation, (\ref{2.1'}) implies
\begin{equation}\label{grthpax}\nabla\Phi=\ve\times\ov\quad\mbox{ in }\Omega,
\end{equation}
where $\ov=\curl\ve$, i.e.,
$$\ov=(\omega_r,\omega_\theta,\omega_z)=\bigl(-\frac{\partial v_\theta}{\partial z},\ \frac{\partial v_r}{\partial z}-\frac{\partial v_z}{\partial r},\ \frac{v_\theta}r+\frac{\partial v_\theta}{\partial r}\bigr).$$
Set $\ov_k=\curl \ue_k$, $\omega(x)=|\ov(x)|$,  $\omega_k(x)=|\ov_k(x)|$.
Since $\Phi\neq \const$ on $V_i$, \eqref{grthpax} implies
$\int_{{\widetilde V}_i}\omega^2\,dx>0$ for every
$i$. Hence, from the weak convergence $\ov_k\rightharpoonup\ov$ in $L^2(\Omega)$  it follows
\begin{lem}
\label{lkr8-ax} {\sl For any $i\in\N$ there exist constants $\e_i>0$,
$\delta_i\in(0,\delta_0)$ and $k'_i\in\N$ such that
$\int\limits_{{\widetilde V}_{i+1}\setminus
\Omega_{\delta_i}}\omega_k^2\,dx>\varepsilon_i$ for all $k\ge
k'_i$.}
\end{lem}

Now we are ready to prove the key estimate.

\begin{lem}
\label{ax-lkr11}{\sl For  any $i\in \N$ there exists $k(i)\in\N$ such
that for every $k\ge k(i)$ and for almost all
$t\in[\frac58t_i,\frac78t_i]$ the inequality
\begin{equation}\label{ax-mec}
\int\limits_{\widetilde S_{ik}(t)}|\nabla\Phi_k|\,dS<\F
t,
\end{equation}
holds with the constant $\F$ independent of  $t,k$ and $i$. }
\end{lem}

\pr Since the proof of this lemma is similar to that of
 Lemma~\ref{lkr11} for the plane case, we comment
only some key steps.

Fix $i\in\mathbb N$. Below we always assume that $k\ge k_i$ (see (\ref{boundary0})\,).
Since we have  removed a neighborhood of the singularity line~$O_z$, we can use the Sobolev embedding theorem in the plane domain~$\D_{r_*}$. In particular, from the
uniform estimate $\|\Phi_k\|_{W^{1,3/2}(\D_{r_*})}\le \const$ we deduce that the~norms $\|\Phi_k\|_{L^6(\D_{r_*})}$ are uniformly bounded.
Consequently, by the H\"older inequality
$\|\Phi_k\nabla\Phi_k\|_{L^{6/5}(\D_{r_*})}\le \const$, and this implies \begin{equation}\label{sest1}
\|\Phi_k\nabla\Phi_k\|_{L^{6/5}(\widetilde\D_{r_*})}\le \const.
\end{equation}

Fix a sufficiently small $\sigma>0$ (the exact value of $\sigma$
will be specified
 below) and take the parameter $\delta_\sigma\in(0,\delta_i]$ (see Lemma~\ref{lkr8-ax}) small enough to
satisfy the following conditions
\begin{equation}\label{ax-lac0.2.1}
\Omega_{\delta_\sigma}\cap \widetilde A^j_i=\Omega_{\delta_\sigma}\cap
\widetilde A^j_{i+1}=\emptyset,\quad j=M',\dots,M,
\end{equation}
\begin{equation}\label{ax-lac0.4}
\int\limits_{\Gamma_{h}}\Phi_k^2\,dS<\sigma^2
\quad\forall h\in(0,\delta_\sigma]\ \ \forall k\ge k'.
\end{equation}
(the last estimate follows from the identity~$\Phi|_{\Gamma_K\cup\dots\cup\Gamma_N}\equiv0$,
the~weak convergence $\Phi_k\rightharpoonup\Phi$ in the space
$W^{1,3/2}(\Omega)$, and~(\ref{sest1})\,).

By a direct calculation, (\ref{NSk}) implies
$$\nabla\Phi_k=-\nu_k\curl\,\ov_k+\ov_k\times\ue_k+\fe_k=-\nu_k\curl\,\ov_k+\ov_k\times\ue_k+\frac{\lambda_k\nu_k^2}{\nu^2}\,\curl\,{\bf b}.$$
By the Stokes theorem, for any $C^1$-smooth closed surface $S\subset\Omega$ and ${\bf g}\in W^{2,2}(\Omega)$ we have
$$\int_S\curl{\bf g}\cdot{\bf n}\,dS=0.$$  So, in particular, 
$$\int_S\nabla\Phi_k\cdot{\bf
n}\,dS=\int_S(\ov_k\times\ue_k)\cdot{\bf n}\,dS.$$

Now, fix a sufficiently small $\varepsilon>0$ (the exact value of
$\e$ will be specified below). For a given sufficiently large
$k\ge k'$ we make a special procedure to find a number $\bar
h_k\in(0,\delta_\sigma)$ such that the estimates
\begin{equation}\label{ax-lac0.5}
\biggr|\int_{\Gamma_{\bar h_k}}\nabla\Phi_k\cdot{\bf
n}\,dS\biggr|\le
2\int_{\Gamma_{\bar
h_k}}|\ue_k|\cdot|\nabla\ue_k|\,dS
<\e,
\end{equation}
\begin{equation}\label{ax-lac0.6}
\int_{\Gamma_{\bar h_k}}|\ue_k|^2\,dS\le C_2(\e)\nu_k^2
\end{equation}
hold, where  $C_2(\e)$ is independent of $k$ and $\sigma$.
This procedure exactly repeats the argument lines of the proof of Lemma~\ref{lkr11}.

The final part of the proof is  identical to that  of Lemma~\ref{lkr11}. We have to
integrate formula (\ref{cle_lap}) (which is valid for the
axially symmetric case as well)  over the three--dimensional domain
$\Omega_{i\bar h_k}(t)$   with $\partial\Omega_{i\bar
h_k}(t)=\Gamma_{\bar h_k}\cup \widetilde S_{ik}(t)$. This means that we have only
 to replace the  curves $S_{ik}(t)$ by the surfaces $\widetilde S_{ik}(t)$ in
the corresponding integrals. $\qed$

\medskip
 Now, we obtain a contradiction by   repeating word by word the proof of Lemma~\ref{lem_Leray_fc} and replacing   the one--dimensional Hausdorff
measure  by the two--dimensional  one, and   the
curves  $S_{ik}(t)$ by the surfaces $\widetilde S_{ik}(t)$ in the
corresponding integrals.

\subsubsection{The case $\esssup\limits_{x\in\Omega}\Phi(x)>\max\limits_{j=0,\dots,N}\widehat p_j$.}
\label{EPcontr-ax-mi}

Assume that (\ref{as-prev-id-ax}) is satisfied and set
$\sigma=\max\limits_{j=0,\dots,N}\widehat p_j$.
Then, as in the proof of Lemma~\ref{mppr1}, we can find
 a compact connected set $F\subset\D\setminus A_\ve$
such that $\diam(F)>0$, $\psi|_{F}=\const$, and $\Phi(F)>\sigma$. Without loss of generality, we may assume that $\sigma<0$ and $\Phi(F)=0$.
Since now it is more difficult to separate $F$ from $\partial\D$ by regular cycles (than in Lemma~\ref{mppr1}), we have  to apply the method of
 Subsection~\ref{Euler-contr-2}. Namely, take a number $r_0>0$ such that
$F\subset D_{r_0}$, the open set $\D_{\varepsilon}=\{(r,z)\in\D:r>\varepsilon\}$ is connected
for every $\e\le r_0$, and conditions~(\ref{ass-ax-bound}) are satisfied.
Then for $\e\in(0,r_0]$ we can consider the behavior of $\Phi$ on the Kronrod trees $T_{\psi,\e}$ corresponding to
the restrictions
$\psi|_{\bar\D_{\e}}$. Denote by $F^\e$ the element of $T_{\psi,\e}$ containing~$F$. Using the
same procedure as in  Subsection~\ref{Euler-contr-2}, we can find $r_*\in(0,r_0]$ such that the following lemma holds.
\begin{lem}\label{Russo}
{\sl There exist $C_j\in[B^{{r_*}}_j,F^{{r_*}}]$, $j=0,\dots, N$, such that
$\Phi(C_j)<0$ and $C\cap L_{{r_*}}=\emptyset$ for all $C\in [C_j,F^{{r_*}}]$. }
\end{lem}

Set $T_\psi=T_{\psi,{r_*}}$, $F^*=F^{{r_*}}$, and $B_j=B^{{r_*}}_j$, i.e., $B_j\in T_{\psi}$
and $B_j\supset \breve\Gamma_j\cap\overline\D_{r_*}$.
As  above, we can change $C_j$ (if necessary) so that Lemma \ref{Russo} takes the following stronger form:
 $$\begin{array}{lcr}
\displaystyle\forall j=0,\dots,M\ \  C_j\in [B_j,F^*],\ \ \Phi(C_j)<0,\\[12pt]
\displaystyle C\cap \partial\D_{r_*}=\emptyset\quad\forall C\in [C_j,F^*],
\end{array}
$$
and
$$
C_0=\dots=C_{M'}.
$$ 
The rest of the procedure of obtaining a~contradiction is done in the same way
as in   Subsection~\ref{max_inside_the_plane_domain}. Namely, we need to take positive numbers $t_i=2^{-i}t_0$, regular cycles $A^j_i\in [C_j,F^*]$ with $\Phi(A^j_i)=-t_i$, and the set $S_{ik}(t)$ with $\Phi_k|_{S_{ik}(t)}\equiv -t$
separating $A^{M'}_i\cup\dots\cup A^N_i$ from $A^{M'}_{i+1}\cup\dots\cup A^N_{i+1}$, etc. The only difference is that we have to integrate  identity (\ref{cle_lap}) over the three--dimensional domains $\Omega_{ik}(t)$ with $\partial \Omega_{ik}(t)=\widetilde S_{ik}(t)$.

\medskip
{\bf Proof of Theorem \ref{kmpTh4.ax}.} Let the hypotheses of Theorem \ref{kmpTh4.ax} are satisfied. Suppose that its assertion fails. Then by Lemma~\ref{lem_Leray_symm} there exist $ \ve, p$ and a sequence $(\ue_k,p_k)$ satisfying  (E-NS-AX). However, in Subsections \ref{EPcontr-axx}--\ref{EPcontr-ax-mi} we have shown that  assumptions (E-NS-AX) lead to a contradiction in all possible cases~(\ref{as-prev1})--(\ref{as-prev-id-ax}). This finishes the proof of Theorem~\ref{kmpTh4.ax}. $\qed$

\begin{lemr}
\label{rem_rot}{\rm Let in Lemma~\ref{lem_Leray_symm}  the data $\fe$ and~$\bf a$ be  axially symmetric without rotation.  If the corresponding assertion of Theorem~\ref{kmpTh4.ax} fails,
then it can be shown (see \cite{kpr_a_arx}) that  conditions~(E-NS-AX) are satisfied
with $\ue_k$  axially symmetric without rotation as well.
But since we have proved that assumptions~(E-NS-AX) lead to a contradiction in the~more general case with possible rotation, we get the validity of both assertions of Theorem~\ref{kmpTh4.ax}. }
\end{lemr}

\begin{lemr}
\label{Russo2}{\rm
It is well know (see~\cite{Lad}) that under hypothesis of Theorems~\ref{kmpTh4.1}, \ref{kmpTh4.ax}, every weak solution $\ue$ of problem~(\ref{NS}) is more regular, i.e, $\ue\in W^{2,2}(\Omega)\cap W^{3,2}_\loc(\Omega)$. }
\end{lemr}

{\small
\section*{Acknowledgements}

$\quad\,\,$The authors are deeply indebted to S.M.~Nedogibchenko
and V.V.~Pukhnachev for valuable discussions.

 The research of M. Korobkov was supported by
the Russian Foundation for Basic Research (project
No.~12-01-00390-a).

The research of K. Pileckas was funded by
the Lithuanian-Swiss cooperation programme
 under the~project agreement No. CH-\u{S}MM-01/01.
}

\end{document}